\documentclass[a4paper,11pt]{article}
\pdfoutput=1 
\usepackage{jcappub} 
\usepackage[T1]{fontenc} 
\usepackage{amsmath}
\usepackage{color}
\usepackage{graphicx}

\usepackage{multirow}

\usepackage{dsfont}

\def\reff@jnl#1{{\rm#1\/}}

\def\aap{\reff@jnl{A\&A}}                
\def\aj{\reff@jnl{AJ}}                  
\def\araa{\reff@jnl{ARA\&A}}            
\def\apj{\reff@jnl{ApJ}}                
\def\apjl{\reff@jnl{ApJ}}               
\def\apjs{\reff@jnl{ApJS}}              
\def\ao{\reff@jnl{Appl.Optics}}         
\def\apss{\reff@jnl{Ap\&SS}}            
\def\aap{\reff@jnl{A\&A}}               
\def\aapr{\reff@jnl{A\&A~Rev.}}         
\def\aaps{\reff@jnl{A\&AS}}             
\def\azh{\reff@jnl{AZh}}                        
\def\baas{\reff@jnl{BAAS}}              
\def\jcap{\reff@jnl{JCAP}}              
\def\jrasc{\reff@jnl{JRASC}}            
\def\memras{\reff@jnl{MmRAS}}           
\def\mnras{\reff@jnl{MNRAS}}            
\def\pra{\reff@jnl{Phys.Rev.A}}         
\def\prb{\reff@jnl{Phys.Rev.B}}         
\def\prc{\reff@jnl{Phys.Rev.C}}         
\def\prd{\reff@jnl{Phys.Rev.D}}         
\def\prl{\reff@jnl{Phys.Rev.Lett}}      
\def\procspie{\reff@jnl{Proc.SPIE}}      
\def\physrep{\reff@jnl{Phys.Rep.}}      
\def\pasp{\reff@jnl{PASP}}              
\def\pasj{\reff@jnl{PASJ}}              
\def\qjras{\reff@jnl{QJRAS}}            
\def\skytel{\reff@jnl{S\&T}}            
\def\solphys{\reff@jnl{Solar~Phys.}}    
\def\sovast{\reff@jnl{Soviet~Ast.}}     
 \def\ssr{\reff@jnl{Space~Sci.Rev.}}    
\def\zap{\reff@jnl{ZAp}}                
\def\nat{\reff@jnl{Nature}}             

\newcommand{\tn}[1]{\mathrm{#1}}
\newcommand{\tr}[1]{\mathrm{#1}}

\newcommand{\inv}{^{\raisebox{.2ex}{$\scriptscriptstyle-1$}}}
\newcommand{\pow}[1]{^{\raisebox{.2ex}{$\scriptscriptstyle{#1}$}}}
\newcommand{\transp}{^{\raisebox{.2ex}{$\scriptscriptstyle{T}$}}}

\newcommand{\Iden}{\mathds{1}}
\newcommand{\Ncov}{\tn{C}_n}

\newcommand{\estm}[1]{\widehat{#1}}
\title{Bandpass mismatch error for satellite CMB experiments II: Correcting for the spurious signal}
\author[a,b]{Ranajoy Banerji,}
\author[a]{Guillaume Patanchon,}
\author[a,c]{Jacques Delabrouille,}
\author[f,g,h,i]{Masashi Hazumi,}
\author[a,d]{Duc Thuong Hoang,}
\author[e]{Hirokazu Ishino,}
\author[e,f]{Tomotake Matsumura,}
\affiliation[a]{Laboratoire Astroparticule et Cosmologie (APC), Universit\'e Paris Diderot, CNRS/IN2P3, CEA/Irfu, Observatoire de Paris, Sorbonne Paris Cit\'e, 10, rue Alice Domon et L\'eonie Duquet, 75205 Paris Cedex 13, France}
\affiliation[b]{Institute of Theoretical Astrophysics, University of Oslo, P.O. Bos 1029 Blindern, N-0315 Oslo, Norway}
\affiliation[c]{D\'epartement d'Astrophysique, CEA Saclay DSM/Irfu,
91191 Gif-sur-Yvette, France}
\affiliation[d]{Department of Space and Aeronautics, University of Science and Technology of Hanoi (USTH), Vietnam Academy of Science and Technology (VAST), 18 Hoang Quoc Viet, Cau Giay District, Hanoi, Vietnam}
\affiliation[e]{Department of Physics, Okayama University,
3-1-1 Tsushimanaka, Kita-ku, Okayama 700-8530, Japan}
\affiliation[f]{Institute of Space and Astronautical Science (ISAS), Japan Aerospace Exploration Agency (JAXA),
Sagamihara, Kanagawa 252-0222, Japan}
\affiliation[g]{High Energy Accelerator Research Organisation (KEK),
Tsukuba, Ibaraki 305-0801, Japan}
\affiliation[h]{Kavli Institute for the Physics and Mathematics of the Universe (Kavli IPMU, WPI), UTIAS, The University of Tokyo, Kashiwa, Chiba 277-8583, Japan}
\affiliation[i]{The Graduate Institute for Advanced Studies (SOKENDAI),
Miura District, Kanagawa 240-0115, Hayama, Japan}
\emailAdd{ranajoyb@astro.uio.no}
\emailAdd{guillaume.patanchon@apc.univ-paris-diderot.fr}
\emailAdd{delabrouille@apc.in2p3.fr}
\emailAdd{hoang@apc.in2p3.fr}
\emailAdd{tomotake.matsumura@ipmu.jp}
\emailAdd{scishino@s.okayama-u.ac.jp}
\emailAdd{masashi.hazumi@kek.jp}
\abstract{Future Cosmic Microwave Background (CMB) satellite missions aim at using the B-mode polarisation signal to measure the tensor-to-scalar ratio $r$ with a sensitivity $\sigma(r)$ of the order of $\leq 10\pow{-3}$. Small uncertainties in the characterisation of instrument properties such as the spectral filters can lead to a leakage of the intensity signal to polarisation and can possibly bias any measurement of a primordial signal. In this paper we discuss methods for avoiding and correcting for the intensity to polarisation leakage due to bandpass mismatch among detector sets. We develop a template fitting map-maker to obtain an unbiased estimate of the leakage signal and subtract it out of the total signal. Using simulations we show how such a method can reduce the bias on the observed B-mode signal by up to $3$ orders of magnitude in power.}
\graphicspath{{Figures/}}
\begin{document}
\maketitle
\flushbottom
\section{Introduction} \label{sec:introduction}

The Cosmic Microwave Background (CMB) is the relic radiation left over from the epoch when electrons and light nuclei from the primordial plasma first combined to form neutral atoms. In the $\Lambda$CDM cosmological scenario, this happened when the Universe was about 380,000 years old. As the properties of the CMB anisotropies depend on the physics that gave rise to initial seeds of structure and also on particle interactions in the primordial plasma, the CMB is a rich source of information about the early Universe. The Planck mission, launched by ESA in 2009 \citep{2011A&A...536A...1P}, delivered to the scientific community full-sky, signal-dominated primordial CMB temperature anisotropy maps on all scales down to about $5'$ resolution, and provided the state-of-the art measurement of CMB polarization of the E-mode type (of even parity) \citep{2014A&A...571A...1P,2016A&A...594A...1P}. These observations put strong constraints on the $\Lambda$CDM cosmological scenario and on its main parameters  \citep{2014A&A...571A..16P}.

However, the observation of one of the remaining key CMB observables, polarization of the B-mode type (of odd parity), has only just barely started. First detections of such B-modes due to the conversion of E-modes from the last scattering surface by intervening matter have been reported \citep{Ade:2013gez,Keisler:2015hfa,Array:2015xqh,Louis:2016ahn,Ade:2017uvt}. However, the very interesting primordial B-modes,  predicted to be generated in the early Universe in a phase of rapid expansion known as cosmic inflation, still escape detection. The expected amplitude of such primordial B-modes is set by the value of a single parameter, $r = \Delta_t^2 / \Delta_s^2$ ($\Delta$ is the amplitude of the fluctuation), which measures the ratio of tensor perturbations to scalar perturbations in the early universe at some reference scale. As of now, combined observations from the BICEP2 and Keck Array experiments and from the Planck mission yield an upper limit of $r<0.06$ at 95\% confidence \citep{Array:2015xqh,Ade:2018gkx}. Proposed future CMB space missions such as the consecutive versions of CORE \citep{Bouchet:2011ck,Delabrouille:2017rct}, LiteBIRD \citep{hazumi/etal:2012,Matsumura:2013aja,ishino/etal:2016}, PICO \citep{Sutin:2018onu,Young:2018aby}, PIXIE \citep{Kogut:2011xw}, PRISM \citep{Andre:2013nfa} or ground-based experiments such as CMB-S4 \citep{Abazajian:2016yjj,Abitbol:2017nao} target a sensitivity $\sigma(r)$ of $10^{-3}$ or better, requiring instrumental sensitivity and control of instrumental systematic effects four orders of magnitude better than for observing CMB temperature anisotropies on large scales.

The measurement of polarization B-modes is a challenging task. Polarization-sensitive detectors used in CMB mapping typically observe a mixture of intensity and polarization, in which the contribution from the sky intensity outshines the primordial polarization B-modes by three to four orders of magnitude, or possibly more. To recover B-mode polarization, intensity signals must be separated from polarization using linear combinations of measurements observing with different polarization angles. Such linear combinations can be obtained either using several observations with the same polarization-sensitive detector re-observing the sky at the exact same location but with different orientations, or using different polarization-sensitive detectors observing with different polarization angles \citep{Natoli:2017sqz}. In this second case, a mismatch of the spectral response between different detectors, with different polarization angles, leads to a leakage of the dominant intensity signal to polarization.

A companion paper develops the data model for the leakage of signal from intensity to polarization due to bandpass mismatch between detectors, and evaluates the level of the bias on the observed B-mode signal from the CMB sky, assuming a similar level of uncertainty in band properties as in Planck HFI \citep{Hoang:2017wwv}. It shows that if the bandpass mismatch is not taken care of by appropriate means, the bias on the B-modes at low multipoles can be of the same order of magnitude as the primordial signal on large scale for a target sensitivity of $\sigma(r) \simeq 10\pow{-3}$. In principle, an ideal and achromatic fast rotating Half Wave Plate (HWP) can be used to mitigate this potential systematic effect, and the development of near-ideal broad-band HWPs is actively pursued. Here, we investigate the effectiveness of an alternative approach to mitigate, in the map-making step, the leakage of signal from intensity to polarization due to bandpass mismatch among a set of detectors when no HWP is used in the observation. We extend the preliminary work done in the context of the CORE systematic effects study outlined in section 6 of \citep{Natoli:2017sqz}. The study of whether it is possible to correct the impact of HWP imperfections is beyond the scope of this paper. 

The rest of this paper is organised as follows. In section~\ref{sec:leakage_model}, we model the bandpass mismatch systematic effect for a set of polarization sensitive detectors and discuss how the mismatched signal projects onto the polarization maps. Section~\ref{sec:mapmaking} is dedicated to the estimation of the noise penalty that occurs when one tries to avoid the bandpass mismatch by making single detector polarisation maps which are combined afterwards, or equivalently by fitting for a pixel-dependent correction term for each detector pair. In section~\ref{sec:correction}, we detail the development of map-making approaches that estimate and correct for the leakage signal to first order. We then demonstrate the effectiveness of the correction method using simulations of data timestreams for a set of detectors and a model of bandpass mismatch leakage due to different foreground contamination of the data stream in section~\ref{sec:simulations_and_results} and discuss the results. In section~\ref{sec:conclusion} we conclude and give an outlook for the correction method for use in future CMB missions and for including various other systematic contamination effects.

\section{Model of the bandpass leakage systematic effect} \label{sec:leakage_model}

Most of the observations of CMB polarisation are obtained using total-power detectors that integrate radiation in wide frequency bands (with typical bandwidth $\Delta \nu/\nu \simeq 0.25-0.35$), along a given linear polarization direction. Focal plane pixels can comprise pairs of associated detectors in which two polarimetric sensors (labelled $a$ and $b$) are sensitive to orthogonal components of the mean squared electric field integrated in the band. Such detector pairs, implemented in practice using polarization sensitive bolometers (PSBs), have been used in the High Frequency Instrument (HFI) of the Planck space mission \citep{Lamarre:2010meg}. Multichroic versions using Transition Edge Sensors (TESs), in which each focal plane pixel comprises six detectors, observing each in one of two orthogonal polarisations and one of three independent frequency bands, are being deployed in ground-based experiments \citep{Suzuki:2012bf,2016SPIE.9914E..17P}, and are foreseen as the baseline for the low frequency telescope in the LiteBIRD space mission \citep{2018arXiv180106987S}.

The frequency band of each detector is defined primarily by a set of stacked filters and on-chip filter banks to the independent radiation, each with a high-pass, low-pass, or band-pass spectral response to the incident radiation. Optical elements such as corrugated horns and lenses also contribute to the selection of photons, as well as does the absorption by the sensors themselves. Overall, each detector $i$ is characterised by a specific spectral response $g_i(\nu)$. Although all the detectors in a given frequency band are designed to have the same spectral response, small variations from detector to detector are unavoidable in practice, and have been observed in the case of Planck HFI, as illustrated in \citep{Planck_2013_IX}, Figs 5-10. As discussed in the companion paper \citep{Hoang:2017wwv}, and detailed below, this induces a bandpass mismatch between the detector bands, which leads to a mismatch between the contributions of the various astrophysical emission components in different detector timestreams after the various detectors have been calibrated on CMB temperature anisotropies, or the dipole. 

\subsection{Band-integrated observations}

For each detector $i$, the observed band-averaged intensity in the direction of sky pixel $p$ can be written as
\begin{equation}
  I_i (p) = \sum_c \int_\nu d \nu \, g_i(\nu) \, I_c(\nu,p),
\end{equation}
where the sum runs over all components $c$ (such as the CMB, galactic synchrotron, free-free, thermal dust emission, ...), $g_i(\nu)$ is the normalised spectral response of detector $i$ such that $\int d \nu g_i(\nu) = 1$, and $I_c(\nu,p)$ is the intensity Stokes parameters of component $c$ at frequency $\nu$ and in pixel $p$. Similar equations hold for Stokes parameters $Q$ and $U$. This model of band-averaged intensity can be recast as
\begin{equation} \label{eq:I_emission_model}
  I_i (p) = \sum_c \gamma_{ic}(p) \, I_c(\nu_0,p),
\end{equation}
where $\gamma_{ic}(p)$ is a bandpass coefficient that depends on the detector $i$ being considered, the component $c$ and, if the frequency scaling of component $c$ is pixel-dependent, also on the pixel of the sky $p$. This formalism is particularly relevant when the emission $I_c(\nu,p)$ of each component can be written as a product of a pixel-dependent template $I_c(\nu_0,p)$ at a reference frequency $\nu_0$, and scaled over frequencies with a single pixel-independent scaling factor $A_c(\nu,\nu_0)$:
\begin{equation} \label{eq:I_emission_scaling}
  I_c(\nu,p) = A_c(\nu,\nu_0) \, I_c(\nu_0,p).
\end{equation}
For all components for which this is the case, $\gamma_{ic}$ does not depend on the pixel $p$, and we have
\begin{equation} \label{eq:gamma_ic}
  \gamma_{ic} = \int_\nu d \nu \, g_i(\nu) \, A_c(\nu,\nu_0).
\end{equation}
To a very good approximation, this is the case for CMB temperature and polarisation anisotropies, and for kinetic and thermal Sunyaev-Zel'dovich (kSZ and tSZ) effects (neglecting the relativistic corrections that modify the latter when interactions with very hot electron gas are considered) \citep{1998ApJ...499....1C,1998ApJ...502....7I}. The emission laws of other astrophysical components, such as Galactic synchrotron, free-free, and thermal dust emission, are only approximately pixel-independent, with varying degree of approximation. The approximation of pixel-independence will be used in section \ref{sec:correction} to develop the method to correct, to first order, polarisation maps from bandpass mismatch leakage. 

To give a specific example, the emission law for galactic thermal dust is well modelled using a modified blackbody of the form
\begin{equation} \label{eq:single-greybody}
  A_{\rm dust}(\nu) \propto \nu^{\beta} B_{\nu}(T_{\rm dust}),
\end{equation}
where $B_{\nu}(T)$ is the Planck blackbody function for temperature $T_{\rm dust}$, and $\beta$ is a spectral index that depends on the chemical composition and structure of dust grains. Fitting such a model on multi-frequency data from the Planck space mission, both $T_{\rm dust}$ and $\beta$ are found to be pixel-dependent, with $T_{\rm dust}$ ranging from about 15$\,$K to about 27$\,$K, and $\beta$ from about 1.2 to about 2.2, with mean values of 1.62 and 19.6\,K for the spectral index and the temperature respectively for intensity maps \citep{Ade:2014zja}. As argued in \citep{2018MNRAS.476.1310M}, this range of temperatures and spectral indices, although compatible with Planck data analysis, may be underestimated by reasons of line-of-sight integration effects.

In the companion paper \citep{Hoang:2017wwv}, the bandpass mismatch error is estimated by assuming the pixel-independence of $\gamma_{ic}$. Although this is sufficient to evaluate the amplitude of the bandpass mismatch leakage to first order, we will relax this assumption in the present paper. This will allow us to estimate the amplitude of residuals at second order, after first order correction of the bandpass leakage systematic effect, in section \ref{sec:simulations_and_results}. We proceed in the rest of the section assuming the pixel-dependence of $A_c$ and hence $\gamma_{ic}$.

The different detectors can be intercalibrated on a strong signal with pixel-independent emission law, such as the CMB dipole and additionally the CMB anisotropies using a blind independent component analysis (ICA) method such as SMICA \citep{2003MNRAS.346.1089D,2008ISTSP...2..735C}, as demonstrated on an analysis of WMAP multi-frequency data \citep{2005MNRAS.364.1185P}. This amounts to dividing equation \ref{eq:I_emission_model} by the $\gamma_i$ coefficient of CMB temperature anisotropies, and redefining the units of intensity. We can then write
\begin{equation} 
  I_i (p) = I_{\rm CMB}(p) + \sum_{c \, \neq \, {\rm CMB}} \gamma_{ic}(p) \, I_c(\nu_0,p).
\end{equation}
Note that, by calibrating over the CMB dipole or temperature anisotropies (and thus expressing the $I$, $Q$ and $U$ Stokes parameters of the observed sky brightness in thermodynamic units of temperature fluctuations of a 2.725\,K blackbody), we normalise the bandpass parameter $\gamma_i$ for the CMB to unity. For the other components $c$, the bandpass parameter $\gamma_{ic}$ encompasses all of the dependence of the signal upon the band spectral response function $g_i(\nu)$ for detector $i$, the spectral emission law of each individual component $c$ and the choice of the reference frequency $\nu_0$. Since unit conversion coefficients are frequency-dependent, the frequency-dependence of $A_c(\nu,\nu_0, p)$ and hence the value of $\gamma_{ic}(p)$ both depend on the choice of units for $I_c(\nu,p)$. 

Using the model of equation \ref{eq:I_emission_model} for the various astrophysical components, the signal for detector $i$ at time index $t$ is
\begin{equation} \label{eq:signal_data_model_1}
  X_{it} = \sum_c \gamma_{ic}(p_t) \left[ I_{c}(p_t) + Q_{c}(p_t) \cos \left( 2\psi_{it} \right) + U_{c}(p_t) \sin \left( 2\psi_{it} \right) \right] + n_{it} ,
\end{equation}
where $p_t$ is the sky pixel pointed to at time $t$, $X$ is the total band integrated noisy signal, $\psi$ is the orientation of the polarisation angle in the local coordinate frame at point $p_t$ on the sky, and $n_{it}$ is the noise of detector $i$ at time $t$.

\subsection{Intensity leakage term due to bandpass mismatch}

For a set of detectors $\{i\}$ in a given frequency band, small variations in the actual spectral response $g_i(\nu)$ between the detectors result in slightly different $\gamma_{ic}$ parameters, which can be modelled as
\begin{equation} \label{eq:factorising_gamma}
    \gamma_{ic}(p) = \gamma_c(p) + \delta\gamma_{ic}(p),
\end{equation}
where $\gamma_c$ is a reference value (e.g., for a reference frequency band) common to all detectors for the component $c$, and $\delta\gamma_{ic}(p)$ is a small detector-dependent and pixel-dependent variation. We can rewrite the signal \ref{eq:signal_data_model_1} of detector $i$, keeping the bandpass variation term coupled to the intensity, as:
\begin{equation} \label{eq:data_model_2}
  X_{it} = \left[ I(p_t) + Q(p_t) \cos \left( 2\psi_{it} \right) + U(p_t) \sin \left( 2\psi_{it} \right) \right] 
  \, + \sum_{c \, \neq \, {\rm CMB}} \delta\gamma_{ic}(p_t) \, I_{c}(p_t) \, + \, n_{it},
\end{equation}
where the term in square brackets is the total sky signal in the ideal frequency band, and the second term is the sum of bandpass leakages from intensity due to astrophysical components. The CMB does not appear in the summation as we assume that the data is perfectly calibrated on a CMB anisotropy signal. As the intensity of the various sky emission sources is at least an order of magnitude larger than their polarised amplitude, we neglect in this model the leakage terms from $Q$ and $U$ Stokes parameters.

We thus decompose our data model into two parts. One containing the (multi-component) sky signal we are interested in measuring, and the other, the spurious leakage term which projects foreground intensity onto the polarisation maps when we solve for $I$, $Q$ and $U$ using a set of different detector measurements in any pixel $p$. This spurious term is what we wish to eliminate so that it does not project onto the final polarisation maps. 

\subsection{Bandpass mismatch error for a detector pair} \label{subsec:bpmm_error_pair} 

In many existing and planned CMB polarisation experiments, detectors are arranged in pairs of orthogonal polarimeters, which we index with $i_a$ and $i_b$ to represent orthogonal detectors in a pair. Measuring conventionally the angle $\psi_{it}$ as that of detector $i_a$, the half difference of the two measurements from  two orthogonal detectors is:

\begin{equation} \label{eq:data_model_dif}
  X_{i_{[a-b]}t} = \left[ Q \cos \left( 2\psi_{it} \right) + U \sin \left( 2\psi_{it} \right) \right] 
  \, + \sum_{c \, \neq \, {\rm CMB}} \Delta\gamma_{ic} \, I_{c} \, + \, n_{i_{[a-b]}t},
\end{equation}
where $\Delta\gamma_{ic} = (\gamma_{i_ac} - \gamma_{i_bc})/2$ is the bandpass mismatch coefficient between detectors $i_a$ and $i_b$ for component $c$. We have dropped the explicit pointing notation $p_t$ since we are in time domain. The subscript ${i_{[a-b]}t}$ conventionally indicates that the set of observations $X_{i_{[a-b]}t}$ is obtained from the half difference of $X_{i_at}$ and $X_{i_bt}$. The first term on the right hand side of equation \ref{eq:data_model_dif} is the useful polarisation signal, the second term is the spurious bandpass mismatch contamination of polarization by intensity, and the noise is the half difference of the individual detector noise terms. Such data stream differences are useful to immediately eliminate the intensity term from the equations, and working directly with only polarisation and bandpass mismatch leakage terms.

It is interesting to note that the polarisation signal on the sky does not contribute to any leakage due to a bandpass mismatch, for detector pairs. If we had considered the spurious $Q$ and $U$ terms in equation \ref{eq:data_model_dif}, they would have appeared with a coefficient proportional to $(\delta\gamma_{i_ac} + \delta\gamma_{i_bc})$ and which is $0$. We demonstrate this using simulations in appendix \ref{sec:appendix_a} figure \ref{fig:ideal_noiseless_spectra}.

\subsection{Bandpass mismatch error in Stokes parameters maps}

Assume now that a set of $2\mathcal{N}_i$ detectors, arranged in orthogonal pairs $\{i_a, i_b\}$, is used to map sky intensity and polarisation in a given frequency channel. Starting from the signal data model in equation \ref{eq:signal_data_model_1}, we can rewrite it in matrix-vector format, as
\begin{equation}
    X = \tn{P} S + n,
\end{equation}
where $X$ is the time-ordered vector of observations of all the $2\mathcal{N}_i$ detectors, $S = [I,Q,U]^T$ is the vector of Stokes parameters in pixel space, $\tn{P}$ is a sparse `pointing matrix' constituted of as many rows as there are observations and as many columns as pixels, for which each non-zero element in a row is of the form $[1, \cos ( 2\psi_{it} ), \sin ( 2\psi_{it} )]$, and $n$ is the noise. The best least square error estimate of the Stokes parameters $S$ is
\begin{equation}  \label{eq:GLS-inversion}
  \widehat S =  [\tn{P}^T \tn{C}_n^{-1} \tn{P}]^{-1} \tn{P}^T \tn{C}_n^{-1} X
\end{equation}
where $\tn{C}_n$ is the noise covariance matrix. If errors between the measurements are uncorrelated and all have the same variance, $\tn{C}_n$ is proportional to an identity matrix, and the solution becomes
\begin{equation} \label{eq:pseudo-inversion}
  \widehat S =  [\tn{P}^T  \tn{P}]^{-1} \tn{P}^T  X.
\end{equation}

If the observations are made with sets of `optimised configurations' of polarimeters with orientations evenly spread in $[0,\pi]$, as often is the case, the map-domain noise covariance matrix of all three Stokes parameters, $[\tn{P}^T  \tn{P}]\inv$, is diagonal \citep{1999A&AS..135..579C}, and the $Q$ and $U$ Stokes parameters in any pixel $p$ are estimated as
\begin{equation}
  \estm Q(p) \propto \sum_{it} \cos \left( 2\psi_{it} \right) X_{it}
\end{equation}
and 
\begin{equation}
  \estm U(p) \propto \sum_{it} \sin \left( 2\psi_{it} \right) X_{it}.
\end{equation}
where the sum over the time index is done for those corresponding to pixel $p$.

When all detectors are arranged in orthogonal pairs, taking into account that $\cos \left( 2\psi_{i_at} \right) = \allowbreak -\cos \left( 2\psi_{i_bt} \right)$ and $\sin \left( 2\psi_{i_at} \right) = -\sin \left( 2\psi_{i_bt} \right)$, the spurious bandpass leakage term in \ref{eq:data_model_dif} projects on the $Q$ and $U$ maps as errors $\delta \estm Q$ and $\delta \estm U$.
\begin{subequations} \label{eq:bandpass_leakage_maps}
  \begin{align}
      \delta \estm{Q}(p) = \frac{1}{\mathcal{N}_{\tr{hits}}(p)}\sum_{c} I_{c}(p) \sum_{i \in \{i_a\}} \Delta \gamma_{ic}(p) \, \sum_t{\cos \left( 2\psi_{it} \right)}, \\
    \delta \estm{U}(p) = \frac{1}{\mathcal{N}_{\tr{hits}}(p)}\sum_{c} I_{c}(p) \sum_{i \in \{i_a\}} \Delta \gamma_{ic}(p) \, \sum_t{\sin \left( 2\psi_{it} \right)},
  \end{align}
\end{subequations}
where the sum over $i$ is over all detector labelled $a$ in the pairs used for making the polarisation maps, the sum over $t$ is for all observations of pixel $p$ with detector pair $i$ at different times, and $\mathcal{N}_{\tr{hits}}(p)$ is the total number of observations by detector pairs in pixel $p$. The leakage in either map is proportional to the bandpass mismatch parameter $\Delta \gamma$. If we consider one single pair $i$ of orthogonal detectors, the leakage of intensity in the estimated $Q$ map is proportional to $\sum_{t} \cos ( 2\psi_{it} )$, and in the $U$ map is proportional to $\sum_{t} \sin ( 2\psi_{it} )$. As already pointed-out in the companion paper, an observing strategy for which these average terms vanish results in vanishing projection of the intensity leakage, due to bandpass mismatch, onto polarization maps. The bandpass leakage term can significantly contaminate B-mode polarisation maps when these maps are reconstructed from the data streams as described above, using data from a set of detectors with mismatched frequency bands. In the next section, we discuss ways to avoid projecting the bandpass mismatch term onto Stokes parameters maps, and the penalty in terms of noise level in the case where the measurements are not obtained with evenly spread polarimeter angles.

\section{Map-making accounting for bandpass mismatch} \label{sec:mapmaking}

\subsection{Single detector map-making} \label{single_detector_mm}

In principle, it is possible to completely avoid bandpass mismatch contamination by making single detector maps of $I$, $Q$, and $U$, and then combining these single detector maps into multi-detector maps by simple averaging the polarisation maps (possibly with pixel-dependent coefficients between the maps, for noise-weighting as a function of the pixel-dependent noise level). This is possible as long as the set of angles $\psi_{it}$ allows inverting, for each detector $i$ independently, the linear system that connects Stokes parameters to observations.  Observing strategies that make use of an ideal rotating HWP, which would spread the polarimeter angles evenly in $[0,\pi]$ for each sky pixel without generating HWP-specific systematics, would allow for such single-detector map-making with no loss of sensitivity. At the other extreme, an observing strategy with no HWP, and in which some sky pixels are observed always with the same scanning orientation, does not allow for disentangling $I$, $Q$ and $U$ for those pixels, and thus does not allow for single-detector map-making. Moreover, depending on the scanning strategy, in the absence of a rotating HWP, it may happen that a large spread of angles can only be achieved after a long period of observation (e.g. one year), which requires, to avoid other possible systematic effects, the stability of the instrument over such long periods.

We consider the intermediate case of an observation strategy with no HWP, but in which each pixel is observed with different scanning orientations. Specifically, we consider the case where the spacecraft is located at the Sun-Earth L2 Lagrange point and spins around a spin axis which itself precesses around the anti-solar direction. Such a scan strategy has been considered for CORE \citep{Delabrouille:2017rct}, LiteBIRD \citep{Matsumura:2013aja} and PICO \citep{Sutin:2018onu}. The choice of the angle between the spin axis and the line of sight ($\beta$), between the precession axis and the spin axis ($\alpha$), and of the periods of precession and of spinning, has an influence on the sky coverage and on the distributions of scanning angles over the sky. A more complete discussion of the trade-offs and of the impact of the parameters of the scanning strategy on the projection of systematic effects can be found in reference \citep{2017MNRAS.466..425W}.

For a given scanning strategy, single detector maps are obtained by solving, for each detector $j$ and each pixel $p$, a linear system of equations such as \ref{eq:signal_data_model_1} for a set of values of $t$. The linear system of equations linking the observations to the sky Stokes parameters is inverted with equation \ref{eq:GLS-inversion}, or with \ref{eq:pseudo-inversion} when only uncorrelated white noise is present. This generates no leakage of intensity into polarisation.

However, the map-making strategy that consists in doing first single detector polarisation maps, and then combining those to produce maps of $Q$ and $U$ (hence, without contamination of polarisation maps by intensity signal from non-CMB components) is in general sub-optimal in terms of noise projection. Indeed, the noise covariance matrix of the $I$, $Q$ and $U$ Stokes parameters is
\begin{equation} \label{eq:bad_cov_matrix}
  \left[ \tn{P}\transp \Ncov\inv \tn{P} \right]\inv = \frac{\sigma_n^2}{\mathcal{N}_{\rm hits}} \begin{bmatrix}
                                                              1 & \langle\cos 2\psi_{jt} \rangle & \langle\sin 2\psi_{jt}  \rangle  \\
                                                              \langle\cos 2\psi_{jt} \rangle & \langle \cos^2 2\psi_{jt} \rangle & \langle \frac{\sin 4\psi_{jt}}{2}\rangle \\
                                                              \langle\sin 2\psi_{jt}  \rangle & \langle \frac{\sin 4\psi_{jt}}{2}\rangle &  \langle \sin^2 2\psi_{jt}  \rangle \\
                                                            \end{bmatrix}^{-1}.
\end{equation}
where $\sigma_n^2$ is the variance of the noise in the detector timestreams, and $\mathcal{N}_{\rm hits}$ the total number of data points falling in pixel $p$. 
Only when each detector $i$ observes each pixel with sets of at least three independent measurements with angles $\psi_{jt}$ evenly spread in $[0, \pi]$ so that $\langle\cos 2\psi_{jt} \rangle = \langle\sin 2\psi_{jt} \rangle =0$ and $\langle \cos^2 2\psi_{jt} \rangle = \langle \sin^2 2\psi_{jt} \rangle = 1/2$, is the noise covariance matrix of the $I$, $Q$ and $U$ Stokes parameters diagonal with minimal volume of the error box \citep{1999A&AS..135..579C}, so that we have
\begin{equation} \label{eq:good_cov_matrix}
  \left[ \tn{P}\transp \Ncov\inv \tn{P} \right]\inv = \frac{\sigma_n^2}{\mathcal{N}_{\rm hits}} \begin{bmatrix}
                                                              1 & 0 & 0 \\
                                                              0 & 2 & 0 \\
                                                              0 & 0 & 2 \\
                                                            \end{bmatrix}.
\end{equation}
This ideal case can be achieved with an ideal HWP to modulate the polarisation. For a non-ideal HWP the bandpass may depend on the HWP angle, and the map-making would require again appropriate measures to avoid projecting bandpass mismatch errors in a similar way as in equation \ref{eq:data_model_2}. Without a HWP a near-ideal spread of scanning angles can be achieved with a spacecraft that rotates around the telescope line of sight, but some unevenness cannot be avoided for off-axis detectors.

\begin{figure}[h]
  \centering
  \includegraphics[width=\linewidth]{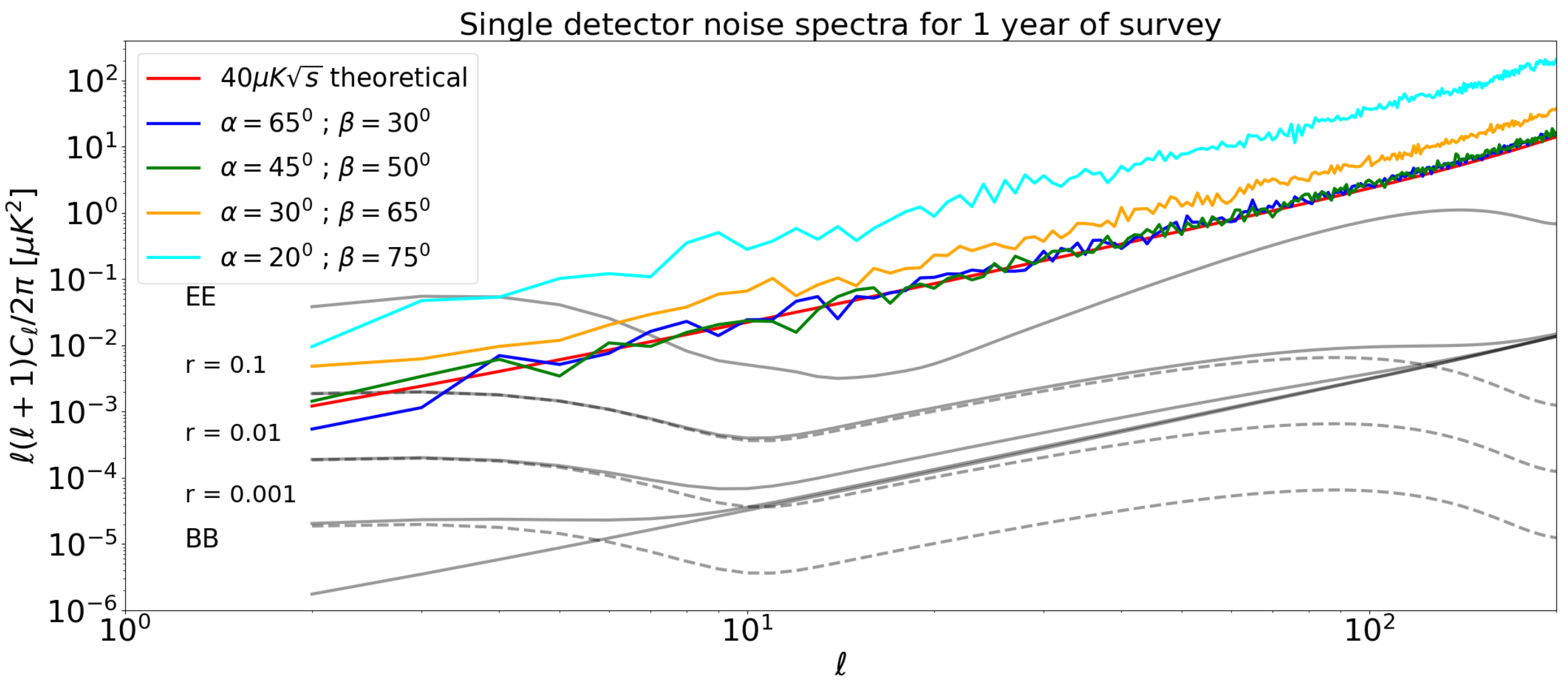}
  \caption[Comparing noise spectra for single detectors with different scan strategies]{Single detector, full-sky noise spectra in polarisation for four scan strategies and for a scan duration of $1$ year are shown for an input noise rms of $40\,\mu$K$\sqrt{s}$. The scan strategies are chosen to demonstrate a wide variety of precession angle ($\alpha$) and spin angle ($\beta$) combinations, keeping $\alpha + \beta = 95^\circ$. The scan strategies have a common spin period of $0.1\,\tr{rpm}$, a precession period of $96.18$ minutes and a sampling rate of $40\,\tr{Hz}$. An ideal and uniform scan would have projected the noise on the theoretical line (solid red). The noise for the different configurations are however more than the theoretical value, increasing with a decrease in $\alpha$. This demonstrates the boosting of the noise spectra as a result of inadequate angular variation and is exacerbated for lower values of precession angle $\alpha$.}
  \label{fig:single_detector_noise_comp}
\end{figure}

In figure \ref{fig:single_detector_noise_comp}, we illustrate, for four different scanning strategies, the BB auto power spectrum of the noise projected onto the polarisation maps. For comparison, we also show the theoretical ideal-case noise level that would be achieved if all of the observation time was evenly spread among all pixels, and scanning angles evenly spread in $[0,\pi]$ for each pixel. Depending on the scan strategy, the noise power spectrum (and hence the white noise variance) after single-detector map-making can be very sub-optimal. This noise increase is due to the large fraction of the sky being sampled poorly leading to a large fraction of the sky having a covariance matrix far from the ideal case, which, for a scan strategy combining rotation and precession, increases with increasing opening angle between the spin axis and the line of sight (spin angle $\beta$). In general, a scan strategy with large precession angle, $\alpha$, and small spin angle, $\beta$, (see figure 1 in reference \citep{Hoang:2017wwv}) results in more evenly spread scanning angles in $[0,\pi]$, and hence smaller noise boosting with single-detector map-making (as compared to map-making using all the detectors with no bandpass mismatch between them). 

\subsection{Leakage subtraction with pair by pair map-making}

In WMAP data analysis, timelines consisting of differences between two detectors are used to make maps of $I$, $Q$ and $U$. Bandpass mismatch between detectors, if neglected, would result in leakages of intensity signal onto polarisation maps. To avoid this contamination, the WMAP team used a map-making strategy that consists in estimating jointly the three Stokes parameters $I$, $Q$, $U$ as well as an extra leakage parameter $M$ \citep{Jarosik:2006ib,Page:2006hz}. This yields the same errors on $Q$ and $U$ maps than single-detector map-making followed by map-averaging. 

Working directly with timelines of differences between two orthogonal detectors $\{i_a,i_b\}$ in a pair, we can write equation \ref{eq:factorising_gamma} for the two detectors in the form
\begin{equation}
  \gamma_{i_ac}(p) = \gamma_{ic}(p) + \delta\gamma_{ic}(p)
\end{equation}
and
\begin{equation}
  \gamma_{i_bc}(p) = \gamma_{ic}(p) - \delta\gamma_{ic}(p),
\end{equation}
where $\gamma_{ic}(p)$ is the average of $\gamma_{i_ac}(p)$ and $\gamma_{i_bc}(p)$ in pixel $p$, and $\delta\gamma_{ic}(p)$ the half-difference of the two. From equation \ref{eq:data_model_2}, noting that $\psi_{it} = \psi_{i_at} = (\psi_{i_bt} \! - \! \pi/2)$, and denoting as $M_i(p)$ the leakage parameter
\begin{equation}
  M_{i}(p) =  \sum_{c \, \neq \, {\rm CMB}} \delta\gamma_{ic} \, I_{c}(p) ,
\end{equation}
for the pair $\{i_a,i_b\}$, we have
\begin{equation}
  X_{i_at}(p) \simeq \left[ I_i(p) + Q(p) \cos \left( 2\psi_{it} \right) + U(p) \sin \left( 2\psi_{it} \right) \right] 
  \, + M_{i}(p) \, + \, n_{i_at},
\end{equation}
and 
\begin{equation}
  X_{i_bt}(p) \simeq \left[ I_i(p) - Q(p) \cos \left( 2\psi_{it} \right) - U(p) \sin \left( 2\psi_{it} \right) \right] 
  \, - M_{i}(p) \, + \, n_{i_bt}.
\end{equation}
Note that the quantity $I_i(p)$ represents the mean intensity for the considered pair, and hence varies from detector pair to detector pair as does the averaged bandpass. This also is the case for $M_i$ which represents the half-difference of intensities $I_{i_a}$ and $I_{i_b}$. This dependence is made explicit with index $i$.\footnote{In principle, the other Stokes parameters, $Q$ and $U$ also depend on the bandpass of each detector, but we neglect this dependence as second order in our map-making model.}

Solving for $[I, Q, U, M]$, for well-balanced detector noise variance, we get
\begin{eqnarray} \label{eq:4x4_map_making}
\begin{bmatrix}
\widehat{I}_i \\ \widehat{Q} \\ \widehat{U} \\ \widehat{M}_i
\end{bmatrix}
= 
\begin{bmatrix}
  1 & 0 & 0 & 0 \\ 
  0 & \langle \cos^2 2\psi_{it} \rangle & \langle \frac{\sin 4\psi_{it}}{2}\rangle &  \langle\cos 2\psi_{it} \rangle\\
  0 &  \langle \frac{\sin 4\psi_{it}}{2} \rangle & \langle \sin^2 2\psi_{it}  \rangle & \langle\sin 2\psi_{it}  \rangle \\
 0 & \langle \cos 2\psi_{it}  \rangle & \langle\sin 2\psi_{it}  \rangle &  1 
 \end{bmatrix}^{-1}
\begin{bmatrix}
    \langle \, X_{i_{[a+b]}t}(p) \, \rangle \\ \langle \, X_{i_{[a-b]}t}(p) \, \cos 2\psi_{it} \, \rangle  \\ \langle \, X_{i_{[a-b]}t}(p) \, \sin 2\psi_{it} \, \rangle \\ \langle \, X_{i_{[a-b]}t}(p) \, \rangle
\end{bmatrix}
\label{matCov}
\end{eqnarray}
where, as in equation \ref{eq:data_model_dif}, we denote as $X_{i_{[a-b]}t}(p)$ the half difference of the two measurements from the two orthogonal detectors at time $t$ in pixel $p$, and  as $X_{i_{[a+b]}t}(p)$ the half sum (quantities that can be computed directly as half-difference and half-sum data streams when the beams of detectors $i_a$ and $i_b$ are co-extensive). The noise covariance matrix of the reconstructed $I$, $Q$, $U$ and $M$ is 
\begin{equation} \label{eq:4x4_cov_matrix}
{\rm Cov}_{4} = \frac{\sigma_n^2}{N_{\rm hits}} \begin{bmatrix}
  1 & 0 & 0 & 0 \\ 
  0 & \langle \cos^2 2\psi_{it} \rangle & \langle \frac{\sin 4\psi_{it}}{2}\rangle &  \langle\cos 2\psi_{it} \rangle\\
  0 &  \langle \frac{\sin 4\psi_{it}}{2} \rangle & \langle \sin^2 2\psi_{it}  \rangle & \langle\sin 2\psi_{it}  \rangle \\
 0 & \langle \cos 2\psi_{it}  \rangle & \langle\sin 2\psi_{it}  \rangle &  1 
 \end{bmatrix}^{-1}.
\end{equation}
We see straightforwardly in equations \ref{eq:4x4_map_making} and \ref{eq:4x4_cov_matrix} that the half-sum timestream contributes only to the measurement of $I$, while the half-difference is used to estimate $Q$, $U$, and the additional leakage term $M$.
From a direct comparison of the noise covariance matrix for $Q$, $U$ and $M$ with that of equation \ref{eq:bad_cov_matrix}, we also notice that this method produces the same $Q$ and $U$ maps as single detector map-making followed by co-averaging the maps obtained by individual detectors. Hence, the same noise penalty incurs on the reconstruction of $Q$ and $U$. In the absence of bandpass leakage that requires estimating $M(p)$ (or equivalently estimating two different intensities for each detector pair, one for detector $i_a$ and one for detector $i_b$), the inverse noise covariance matrix for $Q$ and $U$ would be instead
\begin{equation} \label{eq:3x3_cov_matrix}
{\rm Cov}_{3} =  \frac{\sigma_n^2}{N_{\rm hits}} \begin{bmatrix}
                                                              1 & 0 & 0  \\
                                                              0 & \langle \cos^2 2\psi_{it} \rangle & \langle \frac{\sin 4\psi_{it}}{2}\rangle \\
                                                              0 & \langle \frac{\sin 4\psi_{it}}{2}\rangle &  \langle \sin^2 2\psi_{it}  \rangle \\
                                                            \end{bmatrix}^{-1}.
\end{equation}
In addition to errors on $I$ being decorrelated from those on $Q$ and $U$, the average error on polarisation maps is smaller in that case. 

\begin{figure}[h!]
  \centering
  \includegraphics[width=0.47\textwidth]{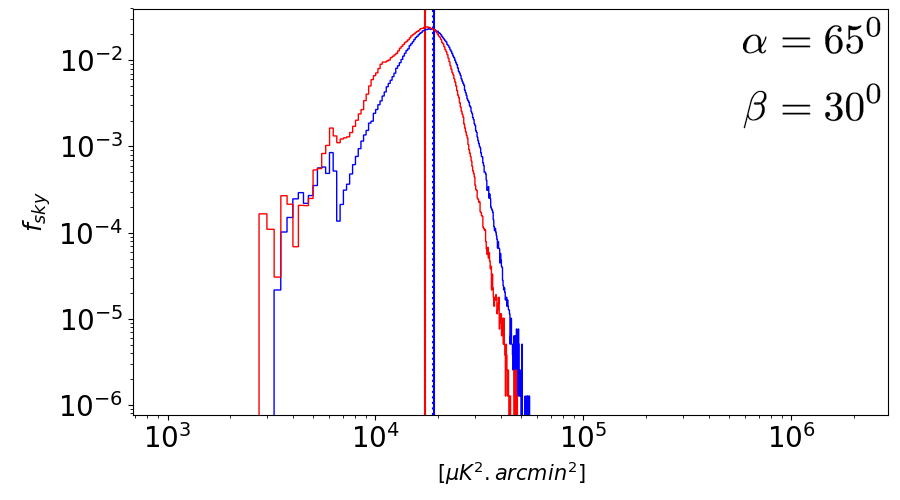}
  \includegraphics[width=0.47\textwidth]{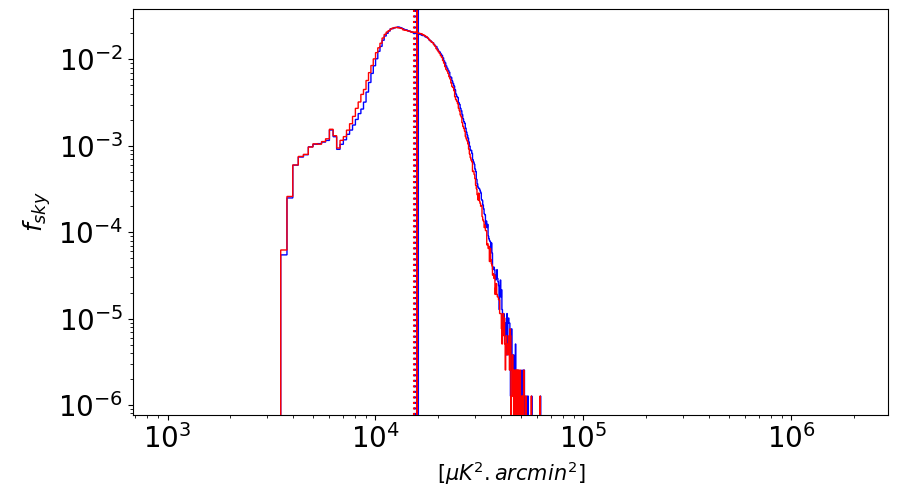}
  \includegraphics[width=0.47\textwidth]{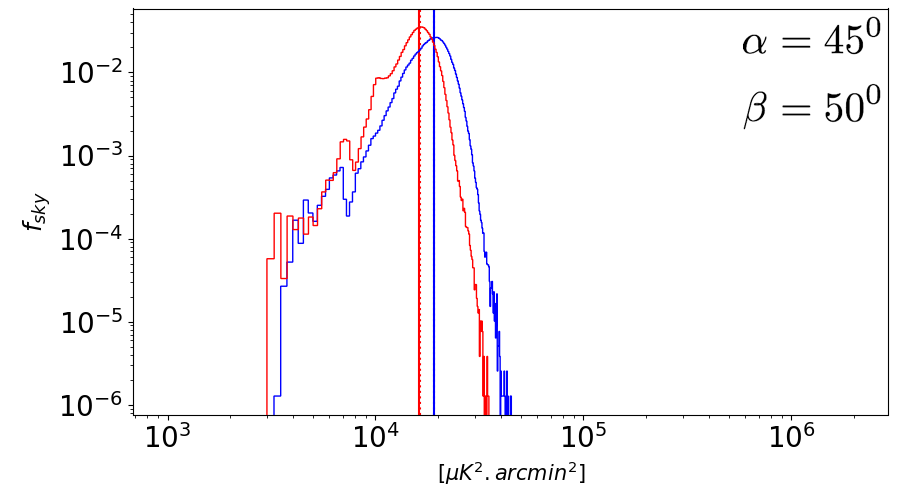}
  \includegraphics[width=0.47\textwidth]{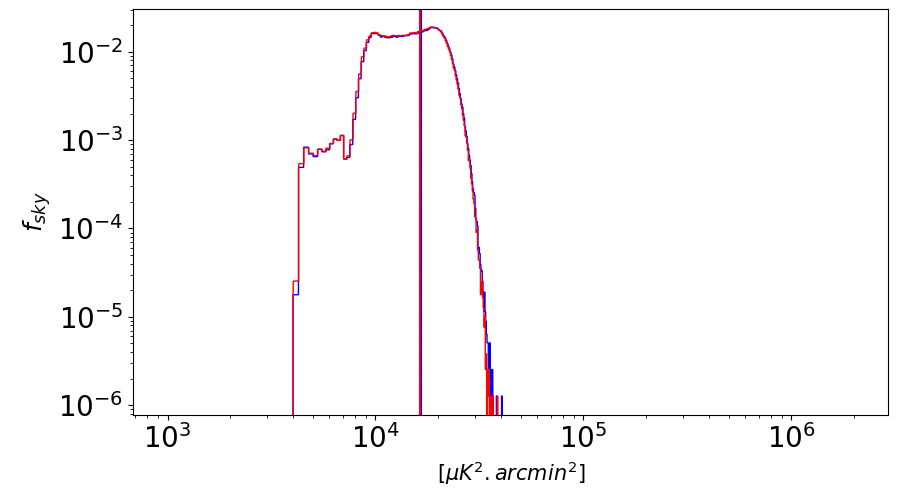}
  \includegraphics[width=0.47\textwidth]{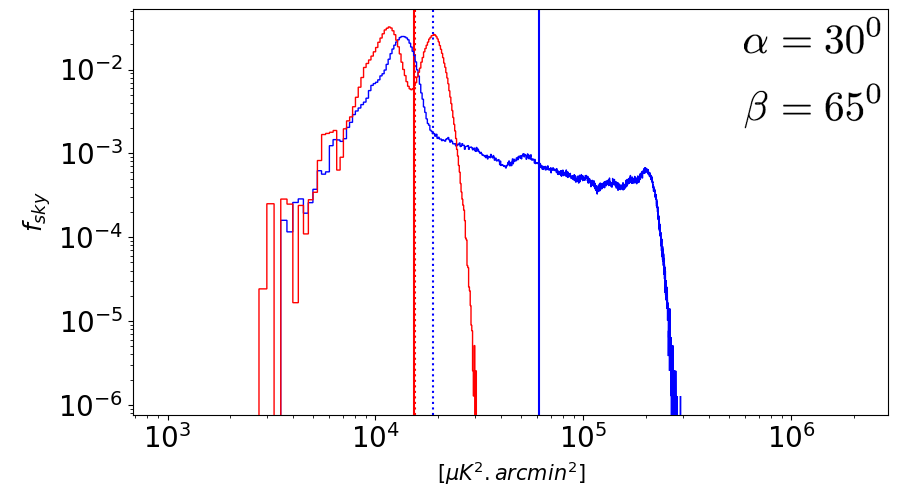}
  \includegraphics[width=0.47\textwidth]{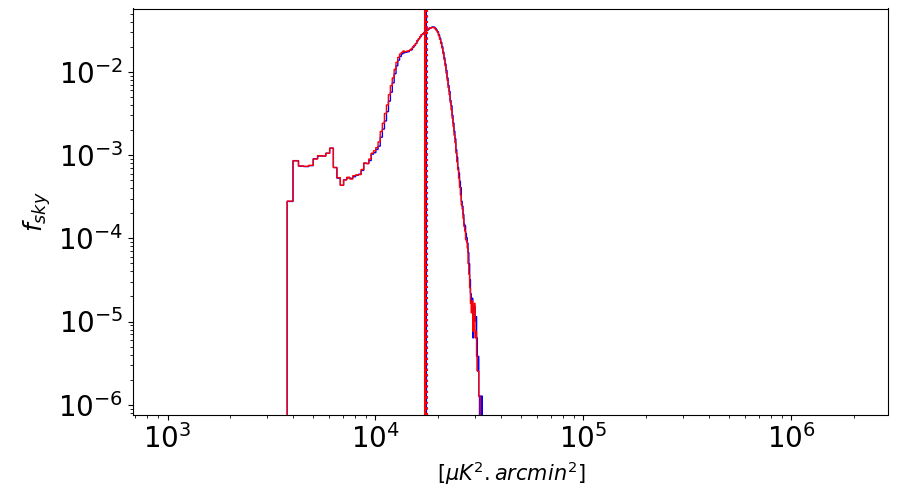}
  \includegraphics[width=0.47\textwidth]{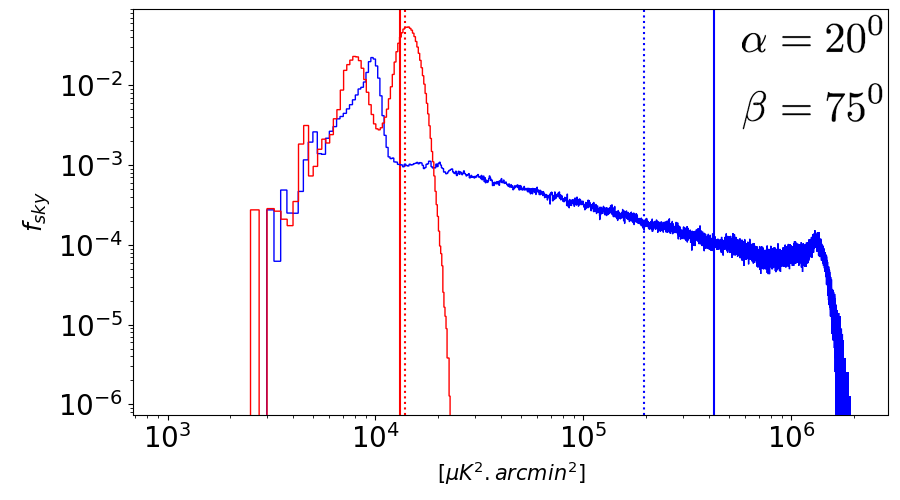}
  \includegraphics[width=0.47\textwidth]{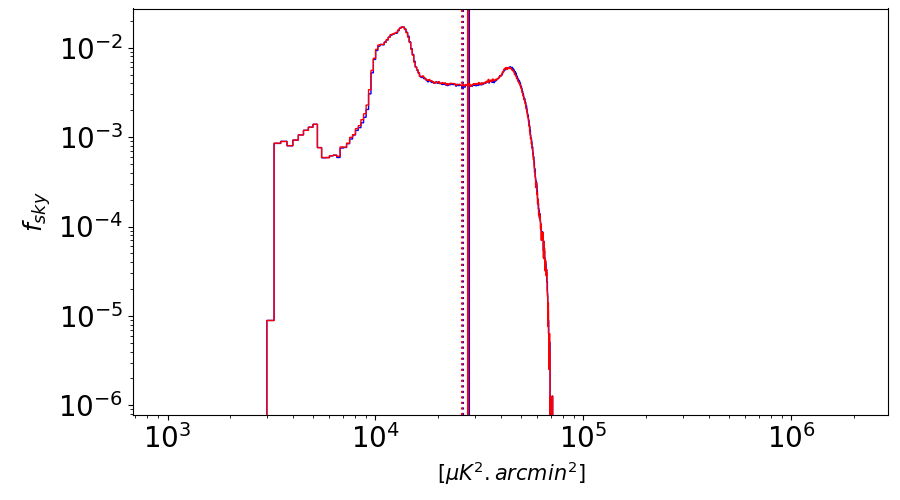}
  \caption[Noise covariance matrix histogram]{Histogram of the noise variance of the $Q$ (left column) and $U$ (right column) polarization component maps, projected in the Ecliptic coordinate system, in the cases where the temperature leakage component is estimated as described by equation \ref{eq:4x4_cov_matrix} (blue), and if we assume no bandpass mismatch error and noise described by equation \ref{eq:3x3_cov_matrix} (red). The corresponding scan strategies for each row is labelled and each of them has a common spin period of $0.1\,\tr{rpm}$, a precession period of $96.18$ minutes and a sampling rate of $40\,\tr{Hz}$. The solid vertical lines correspond to the mean of the histograms of the same colour and the dotted vertical lines their median.}
   \label{fig:histoCov-all}
\end{figure}

Figure \ref{fig:histoCov-all} shows the histograms of the noise variance for the $Q$ and $U$ components (diagonal elements of $\left[ \tn{P}\transp \Ncov\inv \tn{P} \right]$), obtained on simulations with Healpix maps with {\sc Nside} of 256, and for different scanning strategy parameters, after map-making with a pair of detectors, in cases where the leakage component is estimated or neglected (i.e., solving or not for $M(p)$, and hence obtaining noise covariance matrices of either ${\rm Cov}_{4}$ or ${\rm Cov}_{3}$). When solving for $M$, we observe a visible excess of noise for the $Q$ component for all scanning strategies (projected in the Ecliptic coordinate system which allow certain symmetries in the scan strategy). In the cases with $\alpha \! > \! \beta$ and precession period of $96.18$ minutes, the excess is of the order of a few percent, and the distribution for 3 components shows a steeper tail at the high value end. For the strategy with $\alpha \! < \! \beta$, the excess is larger, and for some of the pixels the polarization components are poorly constrained. This degradation of the covariance when the leakage component is estimated is due to sub-optimal coverage of polariser angles in each pixel, in particular in a large region near the ecliptic equator.

We note that the $U$ component is not degraded when solving for $M$, as seen from comparing the two distributions in the figures. This is due to the particular choice of coordinate systems, by reason of symmetries of the scanning strategies in ecliptic coordinates. The study of the global penalty with a full focal plane analysis, and power spectrum estimation with optimal noise weighting of the various map pixels, is postponed to future studies.

\section{First-order bandpass mismatch correction by regression} \label{sec:correction}

We now wish to devise a map-making scheme that corrects for bandpass mismatch while paying the minimum penalty in terms of noise boosting. To do so, we adopt the model developed in equation \ref{eq:data_model_dif}, but drop the dependence of $\Delta\gamma_{ic}$ on the pixel $p$, i.e. we approximate $\gamma_{ic}$ and hence $\Delta\gamma_{ic}$ with its average value on the part of sky of interest for CMB polarisation science. This can be done because (to first order) the frequency scaling of astrophysical components does not vary much from pixel to pixel. For each detector $i$, the pixel-independent term $\gamma_{ic}$ is the bandpass coefficient for an average frequency scaling, given by equation \ref{eq:gamma_ic}. 

Without precise knowledge of the individual detector bands $g_i(\nu)$, the bandpass mismatch coefficients $\Delta\gamma_{ic}$ between detector pairs cannot be computed accurately \emph{a priori}. They must then be estimated from the data themselves. To this effect, we assume that after the mission data is analysed to produce intensity maps, and after component separation for intensity has been performed (see \citep{2009LNP...665..159D} for a review), a template of each of the main sky components $I_{c}(\nu_0,p)$ at some suitable reference frequency in the frequency band of interest (or, up to a redefinition of $\Delta\gamma_{ic}$ that encompasses a factor for the average colour correction for the band and the average emission law of component $c$, a template of the component at any suitable reference frequency $\nu_0$ given by \ref{eq:I_emission_scaling}) will be available at high signal-to-noise ratio. This is, in fact, already the case for the main astrophysical components that have been seen by the Planck and WMAP space missions. Such templates can be produced, for instance, with the Planck Sky Model (PSM) \citep{2013A&A...553A..96D}, or downloaded from the Planck legacy archive at ESA,\footnote{https://pla.esac.esa.int/pla/} or from NASA's Legacy Archive for Microwave Background Data Analysis.\footnote{https://lambda.gsfc.nasa.gov/} The quality of these component intensity templates will still drastically increase with the next CMB space mission.

Assuming then that time-domain templates $I_c$ (derived from the template maps) are available for each of the main components that induce intensity to polarisation leakage by bandpass mismatch, we can recast the data-model adopted from \ref{eq:data_model_dif} with pixel independent bandpass parameters, in a matrix-vector format as,
\begin{equation} \label{eq:data_model_dif_approx2}
X = \tn{P} S + \tn{T} \Delta\gamma + n,
\end{equation}
where the unknowns are $S$, the linear polarisation Stokes parameters $Q$ and $U$ in the frequency band of interest for all pixels in the sky, and $\Delta\gamma$, the bandpass mismatch parameters for each detector pair $\{ i_a,i_b \}$ and for each component $c$. As usual $n$ is the noise. The known quantities are the pointing matrix $\tn{P}$ and the template matrix $\tn{T}$ which is composed of the individual templates $I_c$. 

Solving/inverting equation \ref{eq:data_model_dif_approx2} is a regression problem where we are trying to estimate $S$ and $\Delta\gamma$ given our knowledge of $X$, $\tn{P}$ and $\tn{T}$. Minimizing 
\begin{equation}
  \chi^2 = \left[ X - \tn{P} S - \tn{T} \Delta\gamma \right]^T \tn{C}_n\inv \left[ X - \tn{P} S - \tn{T} \Delta\gamma \right].
\end{equation}
with respect to $S$ and $\Delta\gamma$ yields the maximum likelihood estimators
\begin{equation} \label{eq:direct_estimator_S}
    \widehat{S} = \left[ \tn{P}^T \tn{C}_n\inv \tn{F}_{\tn{T}} \tn{P} \right]\inv \tn{P}^T \tn{C}_n\inv \tn{F}_{\tn{T}} X,
\end{equation}
\begin{equation} \label{eq:direct_estimator_gamma}
  \widehat{\Delta\gamma} = \left[ \tn{T}^T \tn{C}_n\inv \tn{F}_{\tn{P}} \tn{T} \right]\inv \tn{T}^T \tn{C}_n\inv \tn{F}_{\tn{P}} X,
\end{equation}
where
\begin{equation} \label{eq:filter_S}
  \tn{F}_{\tn{P}} = \Iden - \tn{P} \left( \tn{P}^T \tn{C}_n\inv \tn{P} \right)\inv \tn{P}^T \tn{C}_n\inv,
\end{equation}
and
\begin{equation} \label{eq:filter_gamma}
  \tn{F}_{\tn{T}} =  \Iden - \tn{T} \left( \tn{T}^T \tn{C}_n\inv \tn{T} \right)\inv \tn{T}^T \tn{C}_n\inv,
\end{equation}
with $\Iden$ the identity matrix. The $\tn{F}_{\tn{P}}$ and $\tn{F}_{\tn{T}}$ can be thought of as filtering operators which de-project the component of the signal in the space spanned by $\tn{P}$ and $\tn{T}$ respectively. In the case where the noise is well balanced between both detectors and is uncorrelated, $\tn{C}_n \propto \Iden$ and these filtering operators become $\tn{F}_{\tn{P}} = \left[ \Iden - \tn{P} \left( \tn{P}^T \tn{P} \right)\inv \tn{P}^T \right ]$ and $\tn{F}_{\tn{T}} =  \left[\Iden - \tn{T} \left( \tn{T}^T \tn{T} \right)\inv \tn{T}^T \right]$ respectively.

Similar approaches to map-making for CMB observations while solving for nuisance terms in this way have been used in other contexts \citep{Keihanen:2003pu,Poletti:2016xhi,Aghanim:2016yuo}. 

The implementation of the solution of \ref{eq:direct_estimator_S} in the form mentioned above is computationally demanding, consisting of large dense matrices and redundant values. This problem can be bypassed, without requiring sophisticated data distribution and compression algorithms, by performing the map-making in two steps. We first make the unbiased estimation of the set of amplitudes of the templates, $\Delta\gamma$ as in \ref{eq:direct_estimator_gamma}, and then estimate $S$ as:
\begin{equation}
\widehat{S} =  \left[ \tn{P}^T \tn{P} \right]\inv \tn{P}^T \left( X - \tn{T}\widehat{\Delta\gamma} \right).
\end{equation}
We note that in \ref{eq:filter_S} the term $\left[ \tn{P}^T \tn{P} \right]\inv \tn{P}^T$ is simply the map-making equation (written here without inverse noise weighting $\tn{C}_n^{-1}$ as we consider the white-noise case). The operator $\tn{P}$ then re-projects this map back into the time domain. The implementation of the estimation of $\Delta\gamma$ therefore does not require large computational resources, and re-utilises a standard maximum likelihood map-maker. This two step method of estimating the Stokes parameters $S$ is equivalent, for our current purpose, to the direct estimation given by \ref{eq:direct_estimator_S}. The algorithm has been tested on toy models in \ref{sec:appendix_a} and \ref{sec:appendix_b} to demonstrate that it gives mathematically expected results.

In summary, the correction technique for the spurious bandpass leakage signal that is proposed here first estimates the spurious bandpass leakage signal, for each of the contributing sky emission components, and for each orthogonal detector pair. It then subtracts out the estimated spurious signal at the timestream level to de-bias the estimation of the polarisation $Q$ and $U$ Stoke's parameter maps, before re-projecting the corrected data streams onto maps of $Q$ and $U$ in a map-making step.

\section{Simulations and results}  \label{sec:simulations_and_results}

\subsection{Simulated data} \label{subsec:simulated_data}

We evaluate the effectiveness of the correction method described in section \ref{sec:correction} using simulated data streams for $8$ pairs of polarization sensitive detectors arranged in $4$ rows on the focal plane. Two such orthogonal pairs of detectors follow each other on the scan path, are rotated at $45^\circ$ relative to each other, and up to a time-translation, they form an optimal configuration. We consider a scan strategy that is similar to those considered for proposed missions such as LiteBIRD, CORE and PICO. We pick a precession angle $\alpha=45^\circ$, and opening angle for the centre of the focal plane of $\beta=50^\circ$. The spin and precession periods for these particular simulations are $10$ and $96.18$ minutes respectively, which corresponds closely to the LiteBIRD baseline strategy. The observation time is $1$ complete sidereal year. The exact opening angle $\beta + \delta\beta$ of the detectors themselves depend on their location on the focal plane and the value of $\delta\beta$ for the 4 rows of 4 detectors each are $9.875'$, $19.75'$, $29.625'$ and $39.5'$.

We perform our simulation and correction separately for two frequency bands centred near $80\,\tr{GHz}$ (close to the foreground minimum, where dust and synchrotron emission are comparable in amplitude) and $140\,\tr{GHz}$ (one of the preferred frequency for CMB observations), For the $80\,\tr{GHz}$ band we consider separately three cases of components, $1)$ CMB and Thermal Dust, $2)$ CMB and Synchrotron, $3)$ CMB, Thermal Dust and Synchrotron. All $80\,\tr{GHz}$ maps contain the three Stokes parameters $I, Q, U$ and are smoothed by a Gaussian beam of $40'$ FWHM. For the $140\,\tr{GHz}$ band we consider thermal dust as the only foreground component since the contribution from synchrotron is small. These maps are smoothed by a Gaussian beam of $20'$ FWHM. 

To simulate the effect of unequal bandpasses, the foreground components are integrated over top-hat bands that are unique for each detector. A mean bandwidth of $25\%$ of the band central frequency is chosen and the band edges of the individual detectors are sampled from a uniform distribution centred on the mean of the band edges and with a width of $1\%$ of the band centre frequency. This generates the same uncertainty in the bandpass parameter as seen in Planck HFI. In $\mu K$ CMB units, this is equivalent to rescaling the component template maps by the bandpass function $\gamma_{ic}$. The configuration of the detectors on the focal plane as well as the frequency bands are shown in table \ref{tab:det_config}. In the simulated maps, we retain the pixel dependence of the spectral parameters (spectral index and dust temperature for thermal dust, and spectral index for synchrotron) of the foreground components, thus making the bandpass function pixel dependent. All input foreground maps are simulated using the PSM tool using Planck post-flight templates. For the Synchrotron emission we assume a power law scaling, the emission template given by {\tt remazeilles2014}~\cite{Remazeilles:2014mba}, and the spectral index template given by {\tt mamd2008}~\cite{MivilleDeschenes:2008hn} with a mean of $-3.0$ and a standard deviation of $0.06$. We note that these values are slightly lower than those quoted by the S-PASS survey~\cite{Krachmalnicoff:2018imw} conducted over half the sky. For thermal dust we use the {\tt ffp10} model. This is a greybody emission with a mean greybody temperature of $19.4\,\tr{K}$ and standard deviation $1.24\,\tr{K}$, and spectral index with a mean of $1.61$ and standard deviation $0.11$.

\begin{table}[h]
  \begin{tabular}{|| c | c | c | c || c | c | c || c | c | c ||}
    \hline
    Pair & Detector & Pol angle & Position & $\nu_{\tr{min}}$ & $\nu_{\tr{max}}$ & $\nu_{\tr{mid}}$ & $\nu_{\tr{min}}$ & $\nu_{\tr{max}}$ & $\nu_{\tr{mid}}$ \\
    \hline
    & & Degrees & Arc-mins  & \multicolumn{3}{|c||}{$80\,\tr{GHz}$}&\multicolumn{3}{|c||}{$140\,\tr{GHz}$} \\
    \hline\hline
    & $1a$ & 0 & 9.875 & 69.66  & 89.68  & 79.67 & 123.16 & 158.01 & 140.58 \\
    $1$ & $1b$ & 90 & 9.875 & 70.39  & 90.36  & 80.38 & 122.19 & 157.17 & 139.68 \\
    \hline
    & $2a$ & 45 & 9.875 & 69.97  & 90.28  & 80.02 & 122.09 & 158.15 & 140.12 \\
    $2$ & $2b$ & 135 & 9.875 & 70.21  & 90.03   & 80.12 & 121.96 & 157.08 & 139.52 \\
    \hline
    & $3a$ & 0 & 19.75 & 69.7  & 89.97  & 79.84 & 122.62 & 158.04 & 140.33 \\
    $3$ & $3b$ & 90 & 19.75 & 70.11  & 89.84  & 79.97 & 122.62 & 157.02 & 139.82 \\
    \hline
    & $4a$ & 45 & 19.75 & 69.86 & 89.88  & 79.87 & 122.97 & 157.02 & 140.00 \\
    $4$ & $4b$ & 135 & 19.75 & 70.32  & 90.16  & 80.24 & 122.03 & 157.19 & 139.61 \\
    \hline
    & $5a$ & 0 & 29.625 & 69.82  & 90.01  & 79.92 & 121.85 & 157.56 & 139.71 \\
    $5$ & $5b$ & 90 & 29.625 & 69.72  & 89.74  & 79.73 & 123.16 & 157.58 & 140.37 \\
    \hline
    & $6a$ & 45 & 29.625 & 70.01  & 89.81  & 79.91 & 122.84 & 156.91 & 139.88 \\
    $6$ & $6b$ & 135 & 29.625 & 70.36  & 89.70  & 80.03 & 122.37 & 157.54 & 139.95 \\
    \hline
    & $7a$ & 0 & 39.5 & 70.15  & 90.06  & 80.01 & 121.83 & 156.84 & 139.34 \\
    $7$ & $7b$ & 90 & 39.5 & 70.34  & 90.25  & 80.29 & 122.52 & 157.66 & 140.09 \\
    \hline
    & $8a$ & 45 & 39.5 & 69.74  & 89.66  & 79.70 & 122.69 & 158.09 & 140.39 \\
    $8$ & $8b$ & 135 & 39.5 & 69.85  & 90.26  & 80.05 & 122.20 & 158.02 & 140.11 \\
    \hline
  \end{tabular} 
  \caption[Configuration of the focal-plane detectors used in the bandpass mismatch simulation and correction]{The table shows the $16$ detector used in our simulations, their position on the focal plane and their band properties. A positive value of position means it is further away from the spin axis.}
  \label{tab:det_config}
\end{table}

For all sets of simulations, we consider the same noise properties for each detector, a white noise with a rms of $40\,\mu$K$\sqrt{\rm s}$, uncorrelated between detectors.

The thermal dust templates that are used in equations \ref{eq:direct_estimator_S} to \ref{eq:filter_gamma} are derived from a single thermal dust intensity map generated at $353\,\tr{GHz}$ with a delta bandpass. The time-ordered template is generated individually for each detector pair using the pair's pointing and orientation data. Similarly, for synchrotron, the individual templates for each detector pair are derived from a single intensity map generated at $30\,\tr{GHz}$ with a delta bandpass. Both maps are generated using a delta bandpass at their specified frequency. Before generating the time-ordered templates, the foreground template maps are smoothed to the resolution of the frequency band they are correcting for, $20'$ and $40'$ FWHM for $140\,\tr{GHz}$ and $80\,\tr{GHz}$ bands respectively. For a future mission, these intensity maps for the components are expected to be measured at a very high signal to noise ratio but will still have uncertainties. To simulate these uncertainties we overlay a white noise map of $5\,\mu\tr{K.arcmin}$ rms., a reasonable assumption for any future mission of interest.

\subsection{Simulation results}

We carry out the correction on the contaminated time-ordered signal using the iterative algorithm described in section \ref{sec:correction}. We first estimate, using equation \ref{eq:direct_estimator_gamma}, the amplitude of the foreground templates that fit the leakage signal pair by pair. We then subtract out the estimated leakage signal and project the corrected pair differenced signal onto polarization Q and U maps. We note that the correction algorithm approximates the emission laws for the foreground components as uniform on the sky, and hence the bandpass mismatch parameter, $\Delta\gamma$, to be pixel independent. The estimated amplitude of the templates, $\widehat{\Delta\gamma}$, is a global quantity calculated for each detector pair and each foreground component. This allows us a first order correction of the bandpass mismatch signal, with a residual leakage left behind due to the spatial variation of the bandpass mismatch term $\Delta\gamma$. Since we are interested in the polarisation signal away from the galactic plane, we impose a $25\%$ mask on the sky and the template amplitude is estimated only in the valid region.

To observe and analyse the projection of the leakage alone, we subtract out the projected noise at the map level. The map-making process, being linear, allows us this freedom to independently make noise maps. To produce the map of the leakage signal, we first make the uncorrected polarization map for the multi-detector set. The ideal, leakage less, map is that produced from single detectors individually and then averaged. The leakage projection map is then simply the multi detector map with the ideal map subtracted out of it. Similarly, we compute maps of the residual leakage left after correction by subtracting from the corrected multi-detector polarization map the ideal averaged map.
\clearpage
\begin{figure}[h]
  \centering
  \includegraphics[width=0.45\linewidth]{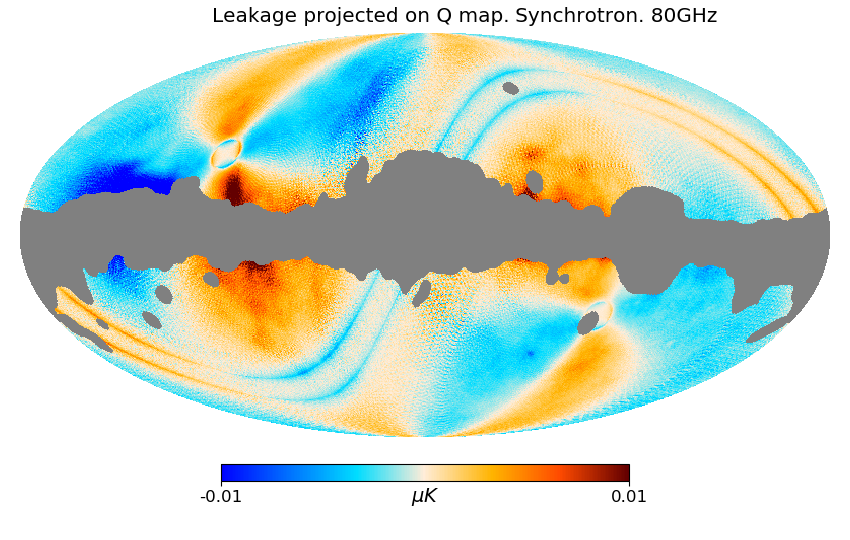}
  \includegraphics[width=0.45\linewidth]{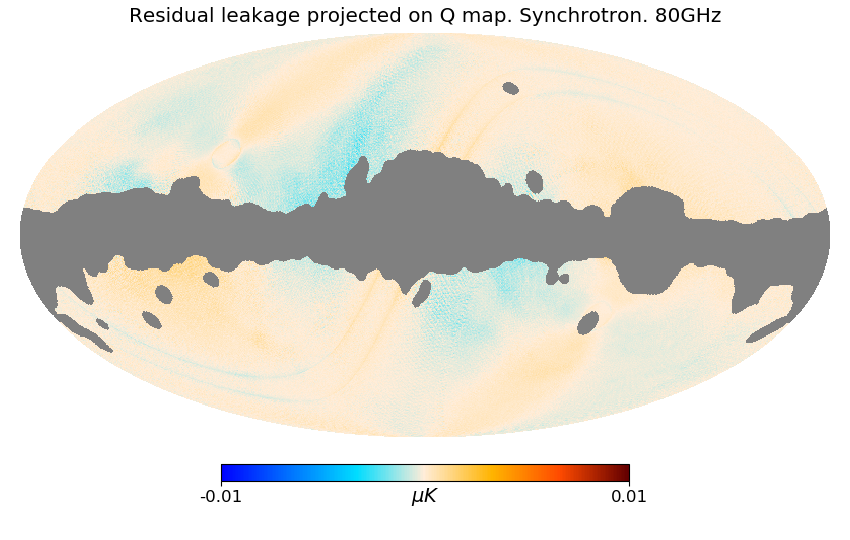}
  \includegraphics[width=0.45\linewidth]{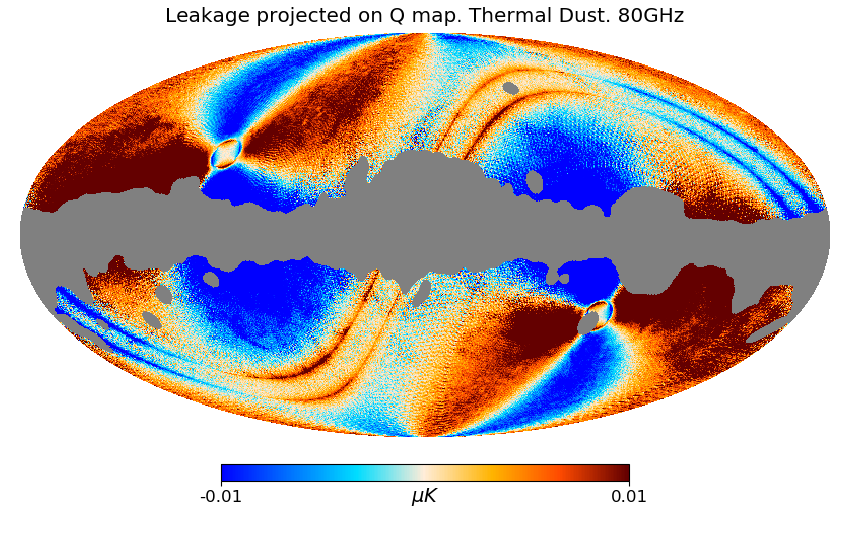}
  \includegraphics[width=0.45\linewidth]{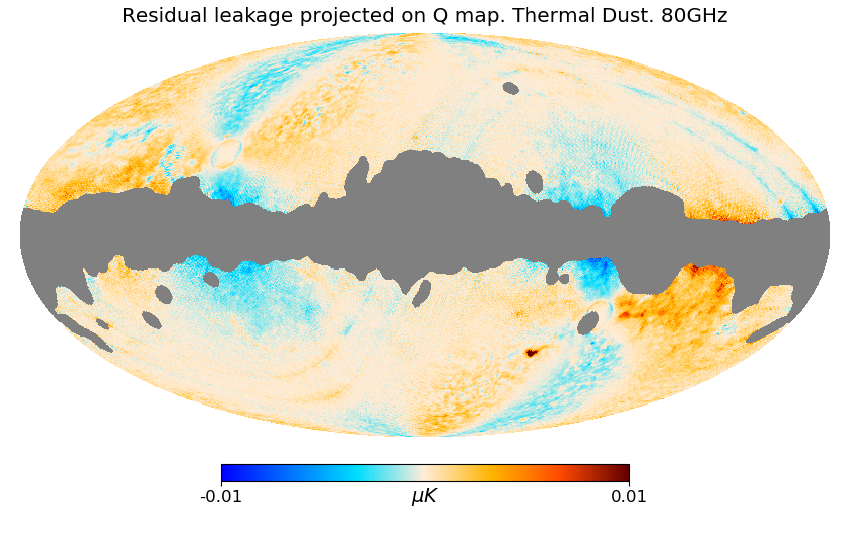}
  \includegraphics[width=0.45\linewidth]{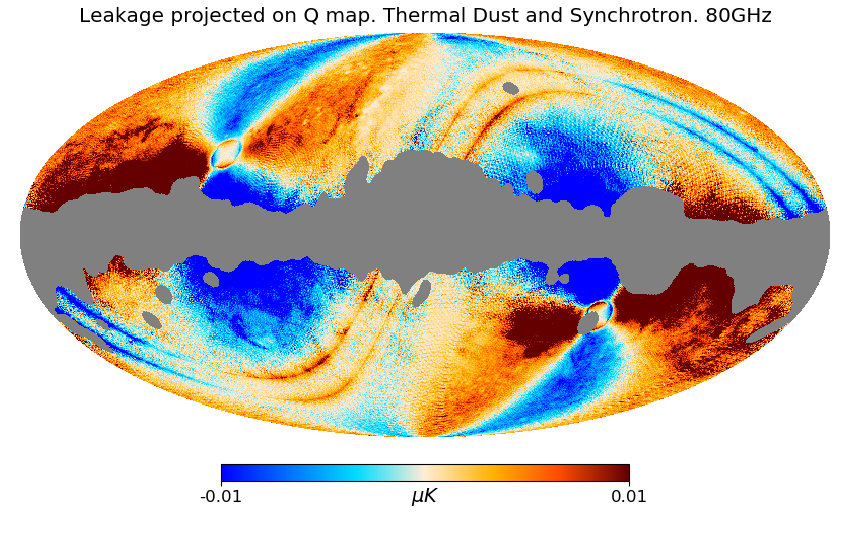}
  \includegraphics[width=0.45\linewidth]{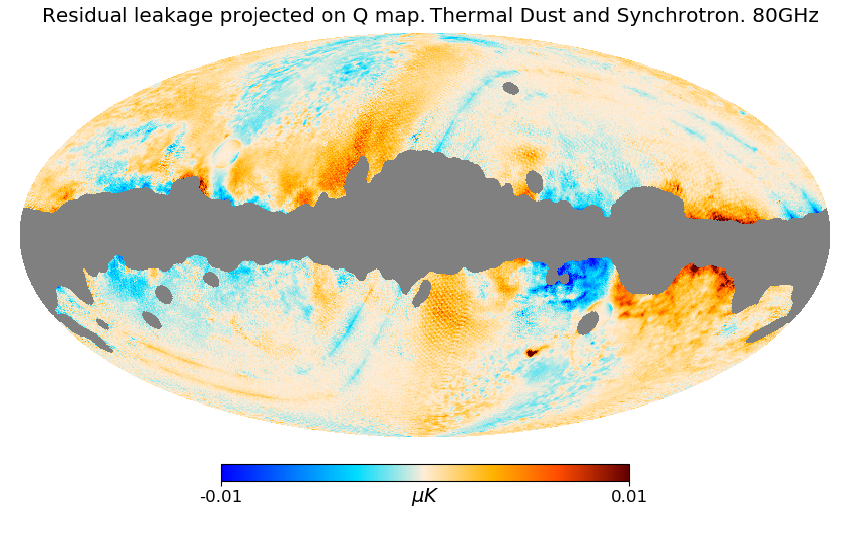}
  \includegraphics[width=0.45\linewidth]{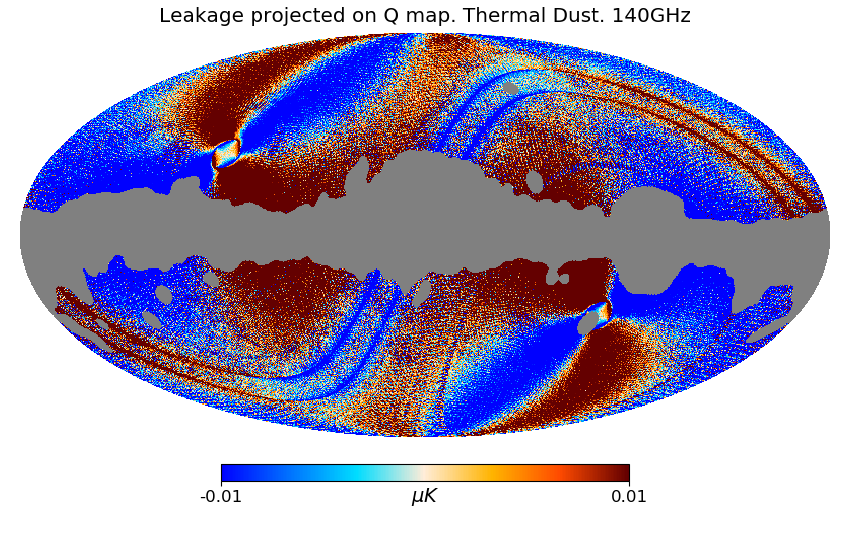}
  \includegraphics[width=0.45\linewidth]{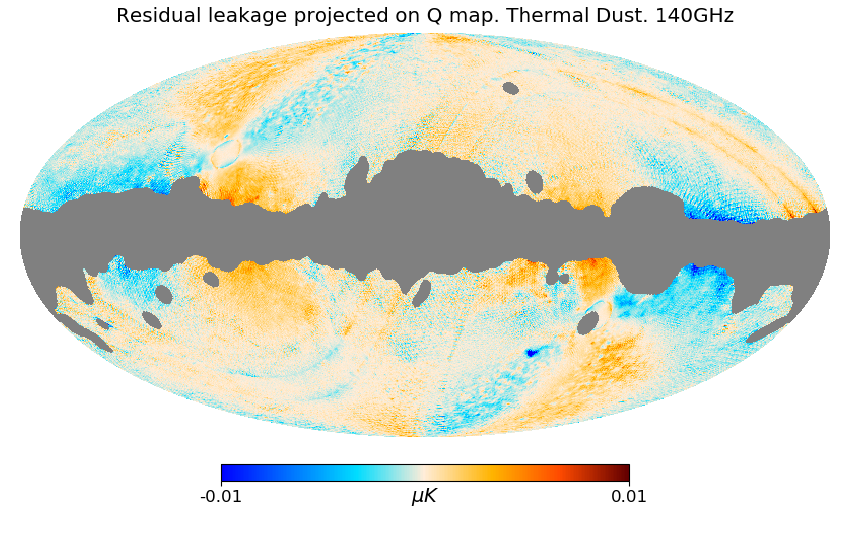}
  \caption{We compare here, for the Stokes Q map, the leakage projected on to the map (left panel) to the residual leakage left after performing the correction (right panel). The first three rows are at $80\,\tr{GHz}$ and the fourth at $140\,\tr{GHz}$. All cases were done for the complete $16$ detector set in the band configuration given in \ref{tab:det_config} and the scanning strategy described in \ref{subsec:simulated_data}. Keeping the same colour scale, we see a significant reduction in the leakage to the polarization maps for all cases. The residual after correction can be attributed to the pixel-independent approximation that the algorithm assumes to estimate the template amplitudes.}
  \label{fig:correction_maps}
\end{figure}

To recapitulate the relevant results from \cite{Hoang:2017wwv}, the spurious signal for each detector pair and for each foreground component is proportional to the foreground intensity signal, the bandpass mismatch parameter and the second order crossing moments $\langle\cos 2\psi \rangle$ and $\langle\sin 2\psi \rangle$. The signal also scales in power inversely with the number of detector pairs. It is also important to mention that the shape of the leakage signal is determined by the foreground intensity signal and the scan strategy. 

In Figure \ref{fig:correction_maps}, we illustrate the magnitude of the correction at the polarization map level by comparing the leakage map to the residual leakage left after the correction has been performed. On the full set of 16 detectors, we see a significant reduction in the leakage on to the polarization maps after correction. The residual leakage is dominated by the effect of the spatial variation of the emission laws of the foreground components. We demonstrate this explicitly in appendix \ref{sec:appendix_c} where we show the effect of noise, spatial variation of spectral parameters and multiple foregrounds. We see that under ideal conditions of no noise and no spatial variations, we are capable of achieving a correction at the level of machine precision even when we use multiple templates.

\begin{figure}[!h]
  \centering
  \includegraphics[width=0.8\linewidth]{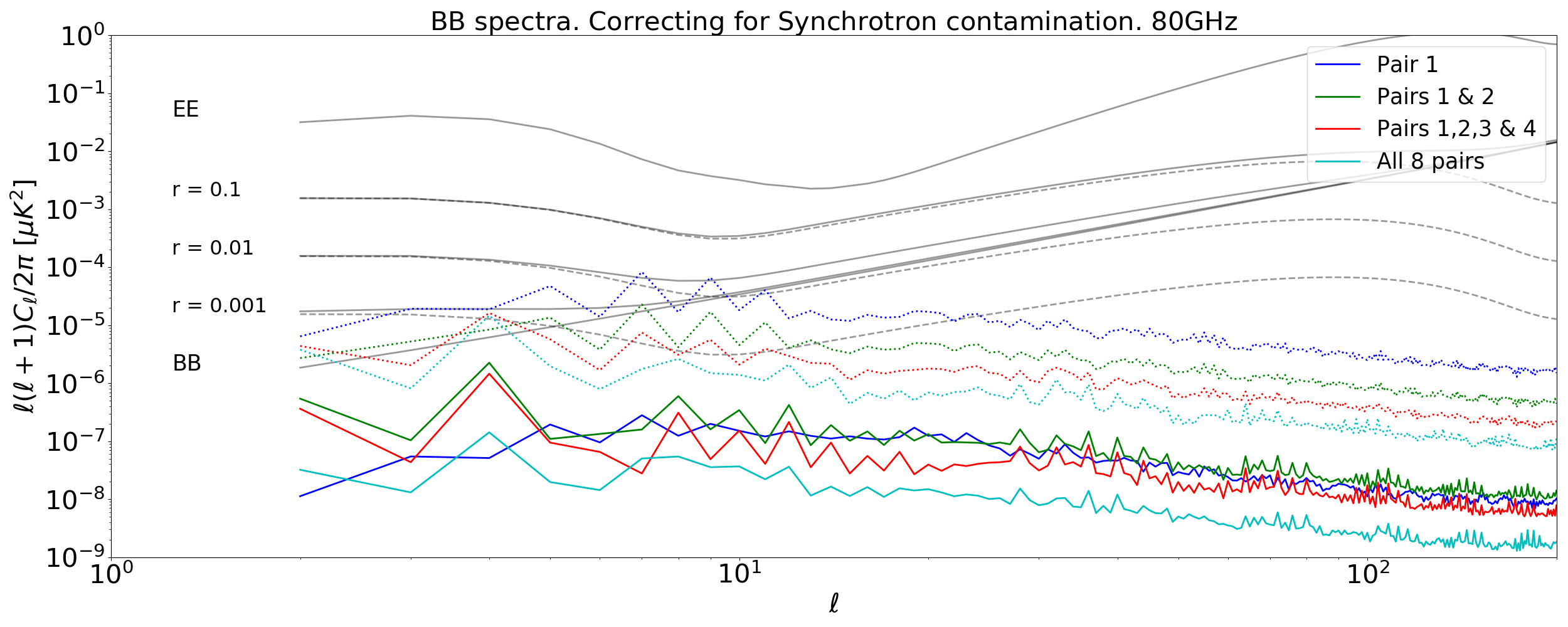}
  \includegraphics[width=0.8\linewidth]{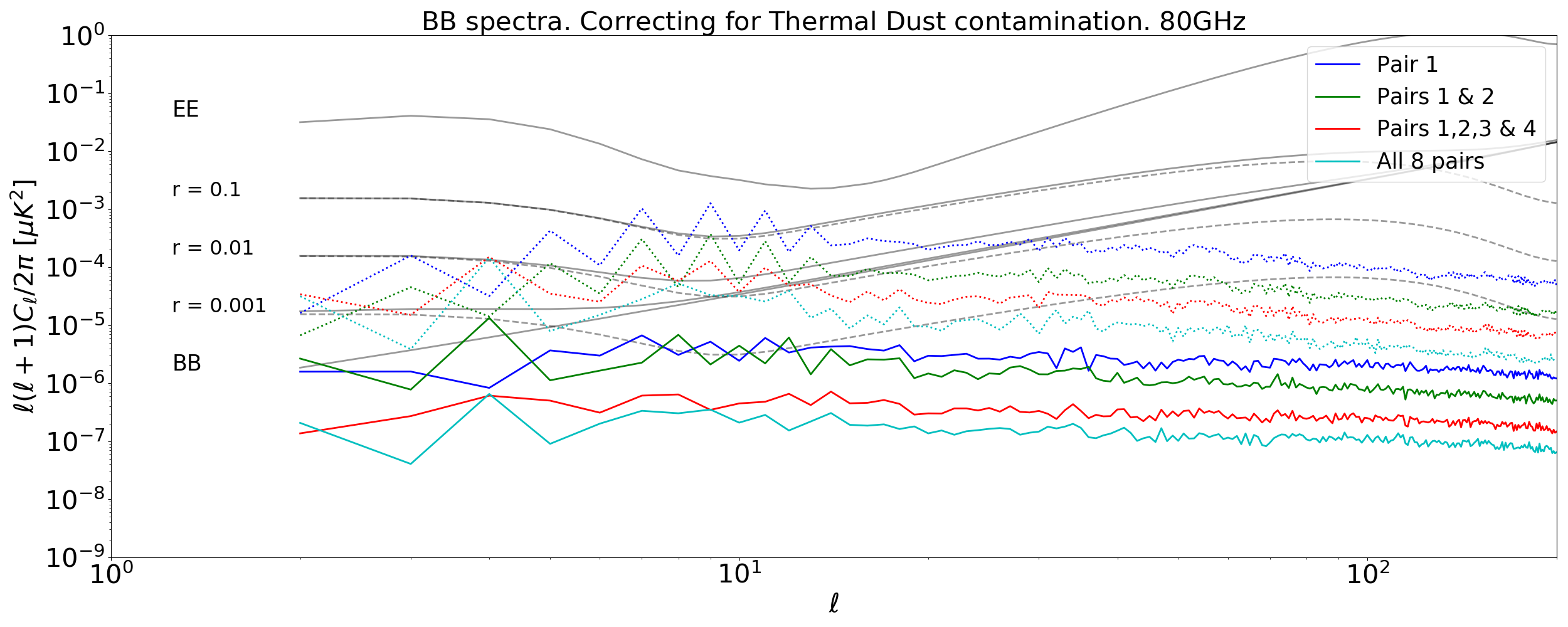}
  \includegraphics[width=0.8\linewidth]{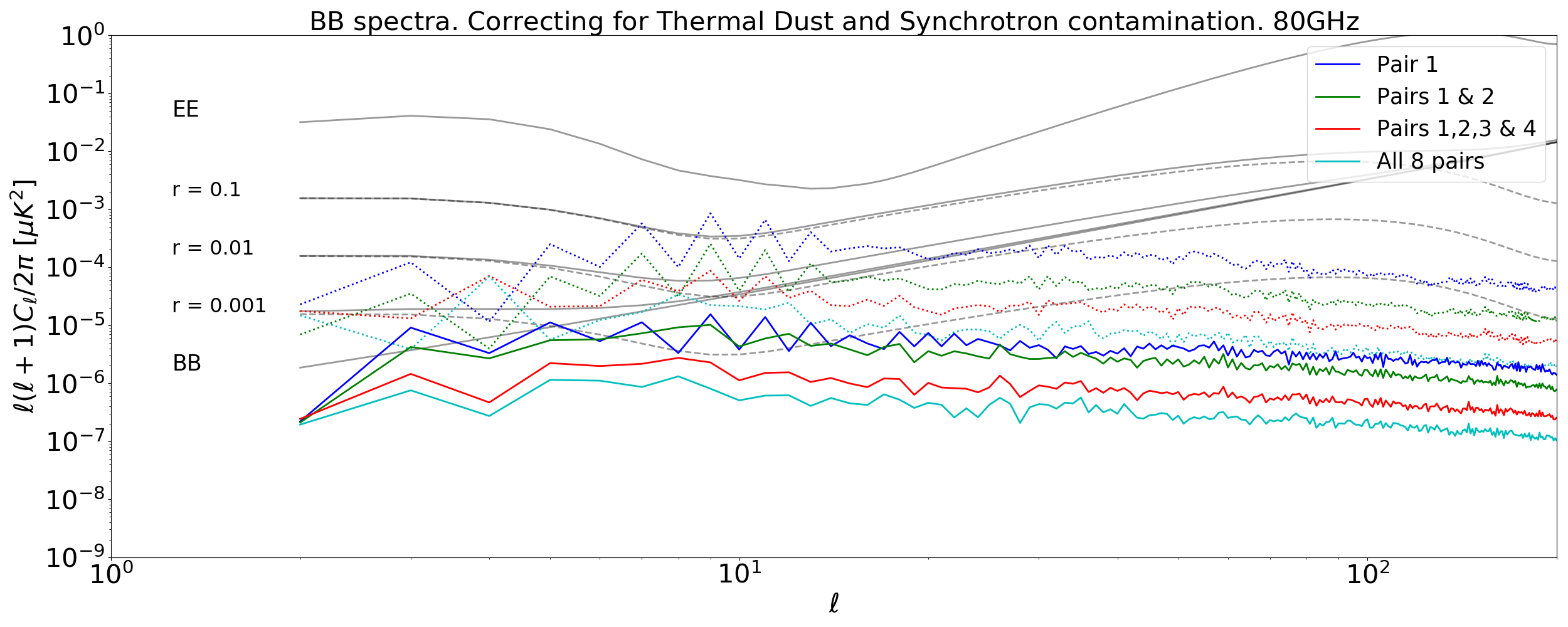}
  \includegraphics[width=0.8\linewidth]{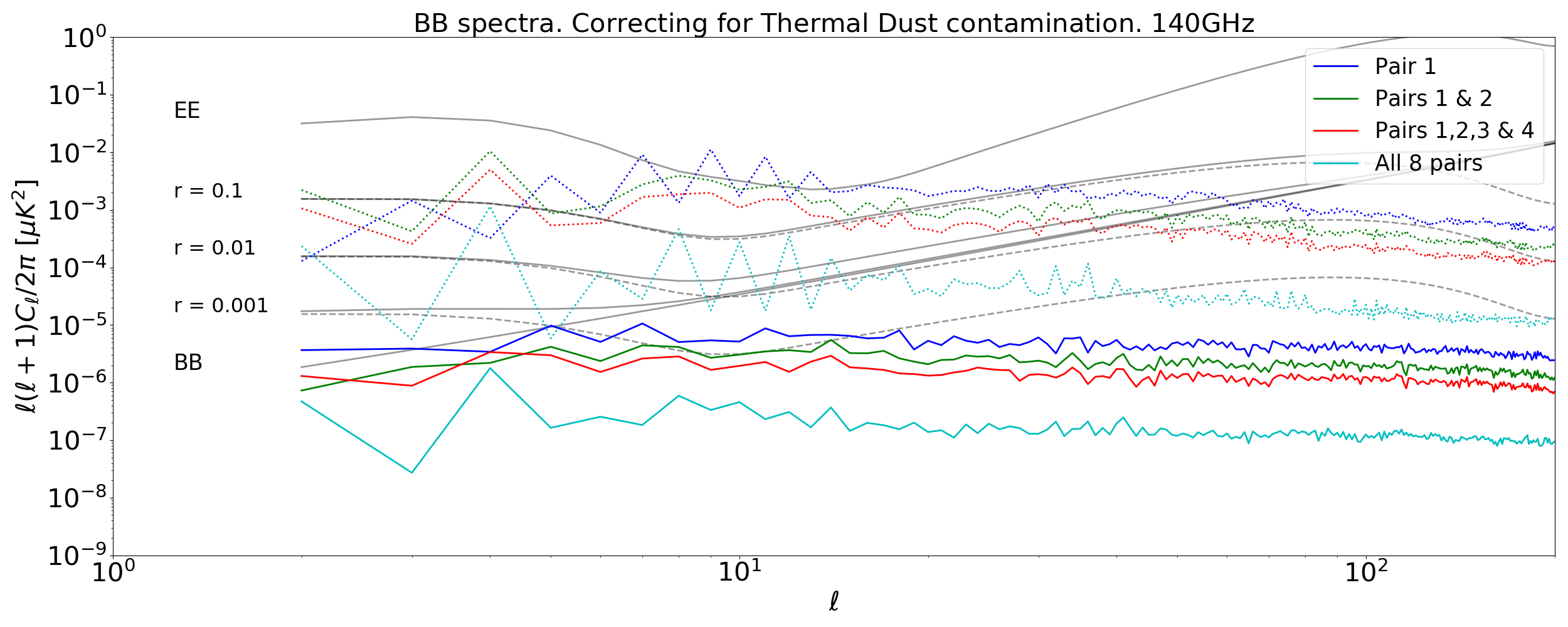}
  \caption{We show the bias produced on the BB power spectrum due to the leakage of signal (dotted lines) and the bias after the correction algorithm is applied (solid lines). As for Figure \ref{fig:correction_maps}, we show the 4 different cases at $80\,\tr{GHz}$ and $140\,\tr{GHz}$ and for different combinations of Synchrotron and Thermal Dust at the input. Having the different combinations of detectors we see the scaling of the residual leakage consistent to that of the leakage signal.}
  \label{fig:correction_spectrum}
\end{figure}

We subsequently quantify the leakage to the BB polarization power spectrum in figure \ref{fig:correction_spectrum}. For this we simply estimate the power spectrum of the masked leakage and the residual leakage maps using Healpix anafast. In all cases, we see a significant improvement, of at least 2 orders of magnitude in power, to the bias in the BB spectrum. The residual leakage scales in a way that is similar to that of the leakage itself. 

\clearpage
\section{Conclusion and Discussion} \label{sec:conclusion}

In the absence of an `ideal' HWP, the bias from the spurious polarisation signal can effectively prevent a detection or reduce the significance of detection for a cosmological signal of $r\sim10^{-3}$. This paper has one driving motivation, to find analysis algorithms which can diminish or effectively cancel completely the leakage of signal from intensity to polarisation due to bandpass mismatch systematics in future CMB space missions that will not employ a continuously rotating HWP. 

In the absence of a continuously rotating HWP, the effect of bandpass mismatch can be avoided altogether if we produce single detector polarisation maps and combine them afterwards, but this solution is suboptimal in terms of noise projection. The spread of polarisation angles in a sky pixel, and hence the noise covariance matrix, depends upon the scanning strategy employed. For intermediate types of scanning strategies envisaged for future missions, the noise covariance matrix only approaches the optimal case when the precession angle is greater than the opening angle of the instrument. In other cases, a sub-optimal spread of polarisation angles boosts the white noise significantly. Moreover, for scans of a shorter period (less than a year), an insufficient spread of angles might prevent inverting the data model altogether. We show that jointly estimating the Stokes parameters $I, Q, U$ along with an additional term for the bandpass leakage, for pair differencing experiments, is equivalent to making single detector maps and averaging them. This incurs the same noise penalty which increases with increasing opening angle of the scan strategy.

To obtain an unbiased estimate of the polarisation $Q$ and $U$ Stokes parameters, without the noise penalty of single detector map-making, we develop a multi-detector map-maker that filters out the intensity to polarisation leakage at the timestream level. We model the leakage term as an additional spurious term proportional to the intensity signal of each foreground component, and the bandpass mismatch parameter for each detector pair. A direct estimation of the unbiased Stokes parameters is computationally intensive and hence we break down the problem in two. First, we estimate the amplitudes of the foreground templates, for each pair of orthogonal detectors. Using this, we subtract out the estimated leakage signal at the timestream level and perform our multi-detector map-making. Using this method we demonstrate that we can reduce the bias on the B mode power by about $3$ orders of magnitude. The residual leakage after correction is primarily due to the non-uniformity of the foreground spectral parameters on the sky. 

The filtering map-making technique demonstrated above is versatile in correcting for any systematic effect that can be modelled as an additional spurious term derived from a template. Effects such as pointing mismatch and beam ellipticity can be modelled in a similar way. We note that for the analysis of real data, refinements to the present algorithm might be required to take into account low-frequency noise, correlated noise among detectors, as well as possible correlations in the bandpass mismatch between similar detector pairs. These extensions of our algorithm are left for future work.

\section{Acknowledgements} 

RB thanks the funding from CNRS-IN2P3 and from the Research Council of Norway through grant 263011 during the development and writing of this paper, and also NERSC for the computing resources. MH is supported in part by JSPS KAKENHI No. JP15H05891. TM was partly supported by World Premier International Research Center Initiative (WPI Initiative), MEXT, Japan, and in part by JSPS KAKENHI No. 18KK0083.

\appendix
\section{Appendix A} \label{sec:appendix_a}

We discuss here the validity and impact of some of the hypotheses we have made in designing our data model and our correction algorithm. In particular, we test the effectiveness of the correction algorithm under different conditions of noise and foreground contamination. 

An important assumption in designing the correction algorithm is that the foreground emission parameters, such as spectral indices, are constant over the entire sky. In practice, these parameters are spatially variable, and the correction algorithm can only achieve a global fit for each parameter. This allows us only to perform a leading order estimation of the leakage and we expect to be left with residues after correction. We construct a few test cases under ideal conditions of spectral parameters to show that our correction algorithm performs at the theoretical limit it is designed for. 

We simulate our toy model skies at the frequencies for detectors $1a$ and $1b$ in table~\ref{tab:det_config} and in the same focal plane configuration. All frequency maps including the template skies, required for constructing our time-ordered templates, are obtained using the PSM tool. For the ideal case, we scale the foreground templates using spatially uniform spectral parameters: a modified black-body model for thermal dust emission with a spectral index $\beta_{\tr{dust}} = 1.62$ and dust temperature $T_{\tr{dust}} = 19.6\,\tr{K}$, and a power law model for synchrotron emission, with spectral index of $\beta_{\tr{sync}} = -3.1$ (in Rayleigh-Jeans brightness temperature units). 

Our first test is designed to evaluate the contribution of the leakage from Intensity to Polarisation as compared to the cross-Polarisation leakage, and the effect of the correction on the leakage from either source. The test is done at the $140\,\tr{GHz}$ band and the input maps contain only CMB and the dominant thermal dust emission. We perform our test with three setups, the first where the input maps contain the Stokes $I$ alone, the second containing Stokes $Q$ and $U$ and the third with all Stokes $I$, $Q$ and $U$ parameters. This allows us to determine exactly the source of each leakage component and the residual leakage component. In all three cases we perform our correction using the template constructed from the $353\,\tr{GHz}$, uniformly scaled, thermal dust intensity map. Here, we de not add  noise to the time stream data nor to the dust template map.

\begin{figure}[!h]
  \centering
  \includegraphics[width=\linewidth]{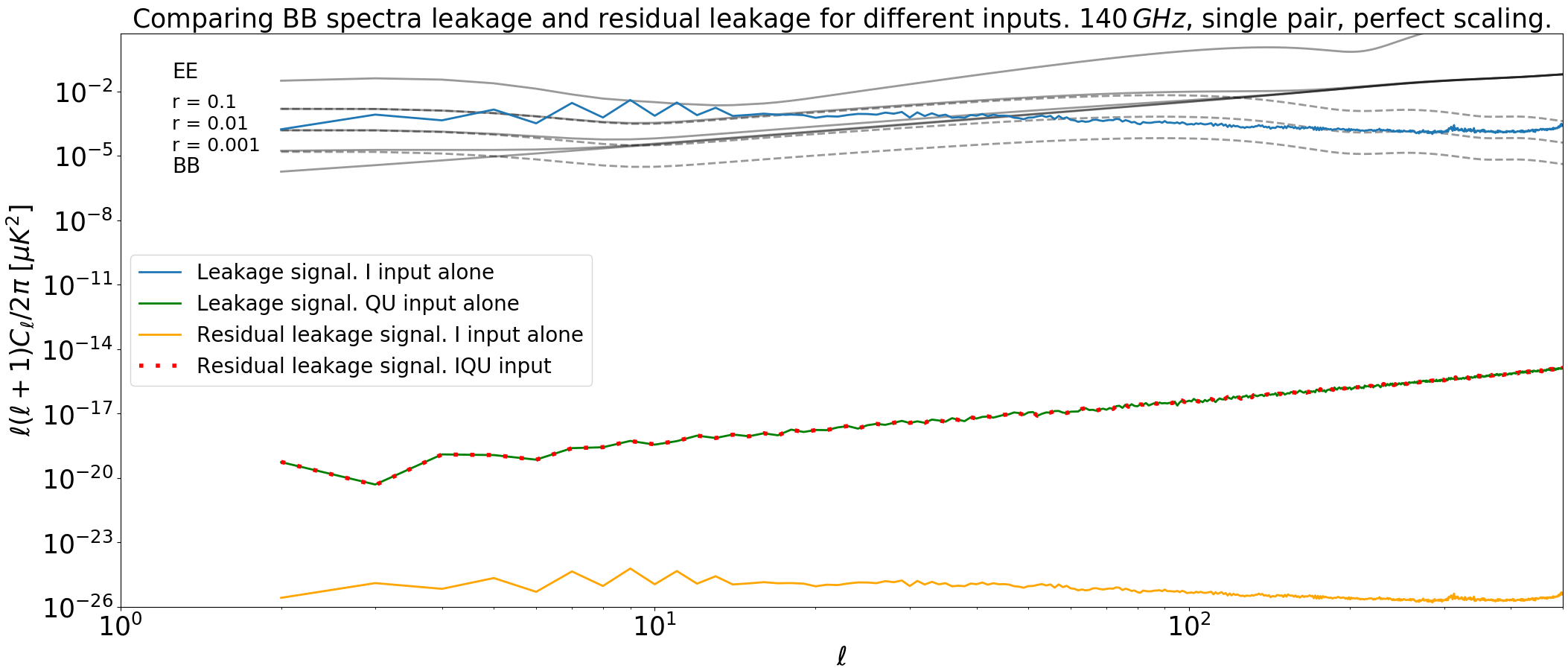}
  \caption{We compare here the intensity to polarisation versus the cross-polar bandpass leakage to the polarisation spectrum (BB), for a single orthogonal detector pair at $140\,\tr{GHz}$, and the only contaminating source being thermal dust. The spectral index and temperature for the thermal dust component is uniform throughout the sky. To demonstrate the leakage we provide two input cases, one with only the intensity map scanned (blue), and one with only the polarisation maps scanned (green), the latter showing an `un-resolved' leakage component. The correction for the intensity to polarisation leakage component is demonstrated using two input cases, one with only the intensity map scanned and one with all intensity and polarisation maps scanned. In the former case, the residual leakage is down at the level of machine precision (orange), and the latter showing a residue at the level of the `un-resolved' polarisation leakage signal since it was unaccounted for in the correction procedure.}
  \label{fig:ideal_noiseless_spectra}
\end{figure}

In figure \ref{fig:ideal_noiseless_spectra} we see a subdominant cross-polar leakage signal in the case where we had just $Q$ and $U$ as input. This can possibly be attributed to the fact that Healpix pixelised polarisation maps suffer from a parallel transport effect of the polarisation angles when the map resolutions are altered or when they are beam smoothed, but they are unlikely to be cross-polar leakage as discussed in section \ref{subsec:bpmm_error_pair}. Let us call this an `un-resolved' source of leakage. From the same figure we see that in the case where we have only Stokes $I$ in the input, the residual leakage after correction can be attributed to machine precision error. This shows us that under ideal noiseless conditions when the sky emission scales uniformly over the sky, the algorithm is capable of performing an exact unbiased estimate of the leakage signal. Since the algorithm corrects only for intensity to polarisation leakage, the `un-resolved' cross-polar leakage is left behind after correction in the case where the input contains all the Stokes parameters.

\section{Appendix B} \label{sec:appendix_b}

Our second test concerns the correction of the leakage when there are more than one mismatched foreground components. For this purpose we build our test case at $80\,\tr{GHz}$ where the intensity of thermal dust and synchrotron are comparable to each other. As before, we do not include noise in the timestream or the template map.

\begin{figure}[!h]
  \centering
  \includegraphics[width=\linewidth]{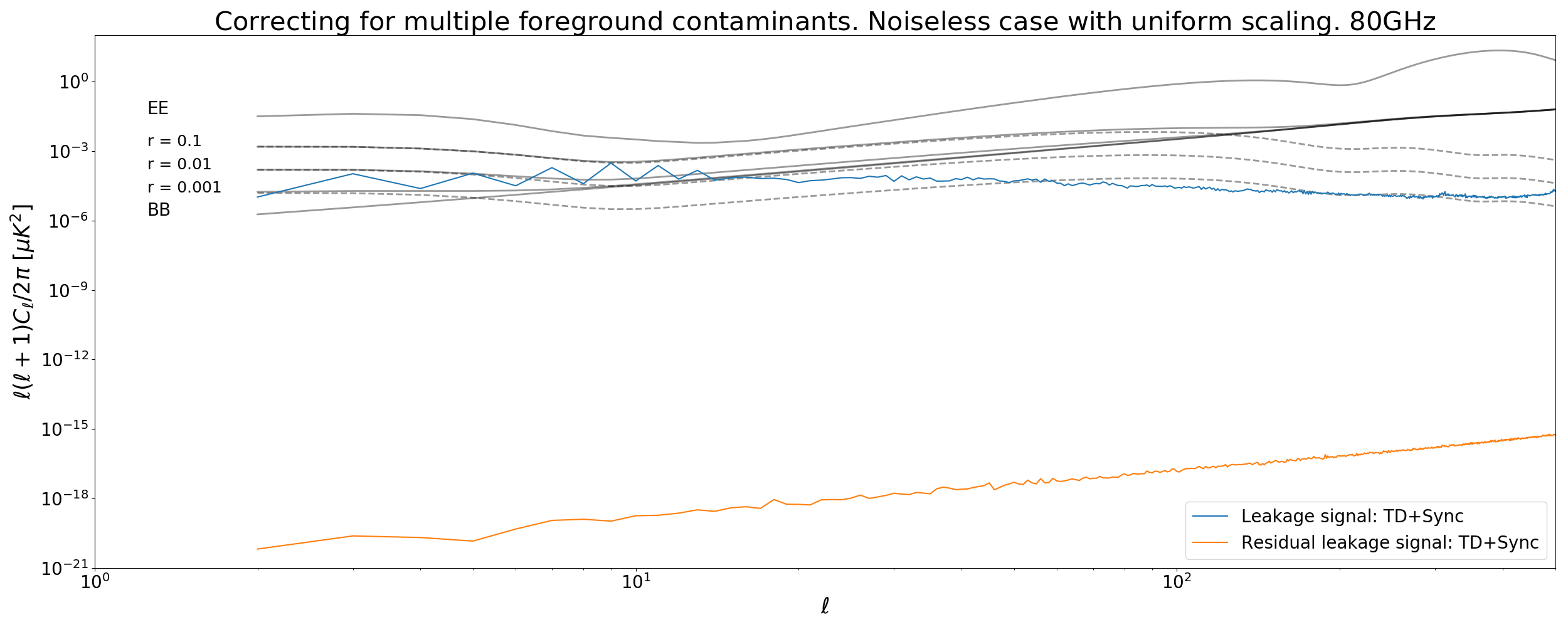}
  \caption{We demonstrate the correction algorithm for multiple input foreground sources (thermal dust and synchrotron) that scale uniformly over the sky and with no input noise in the timestream or on the template maps. In our input we provide only the intensity maps to keep our results unambiguous due to any `un-resolved' polarisation leakage signal. The spectrum of the residue (orange) shows that the multiple contaminants were corrected for upto machine precision.}
  \label{fig:multi_foreground_ideal_correction}
\end{figure}

The leakage signal in figure~\ref{fig:multi_foreground_ideal_correction} contains contribution from both thermal dust and synchrotron. The residual leakage is limited by the spurious cross-polar leakage we have seen in the first test, suggesting that our estimator is unbiased even when more than one leakage component is estimated simultaneously, and there is no cross-talk between components that may arise due to possible degeneracies.

\section{Appendix C} \label{sec:appendix_c}

Our third test case is designed to check whether the correction algorithm is noise limited or limited by the spatial variation of the spectral parameters. We construct two sets of input maps for the same detector pair as used in the previous tests. For simplicity we consider only the CMB and thermal dust component in the $140\,\tr{GHz}$ band. In addition to the uniformly scaled foreground emission maps we construct our second set of sky maps with a spatially varying spectral index and dust temperature, as we have done in section~\ref{sec:simulations_and_results}. The thermal dust template map used in both cases contains a $5\,\mu\tr{K.arcmin}$ white noise. For both sets of input maps we perform five sets of simulations and correction, each with a different white noise realisation of r.m.s $40\,\mu\tr{K}\sqrt{\tr{s}}$ in the timestream.
\begin{figure}[!h]
  \centering
  \includegraphics[width=\linewidth]{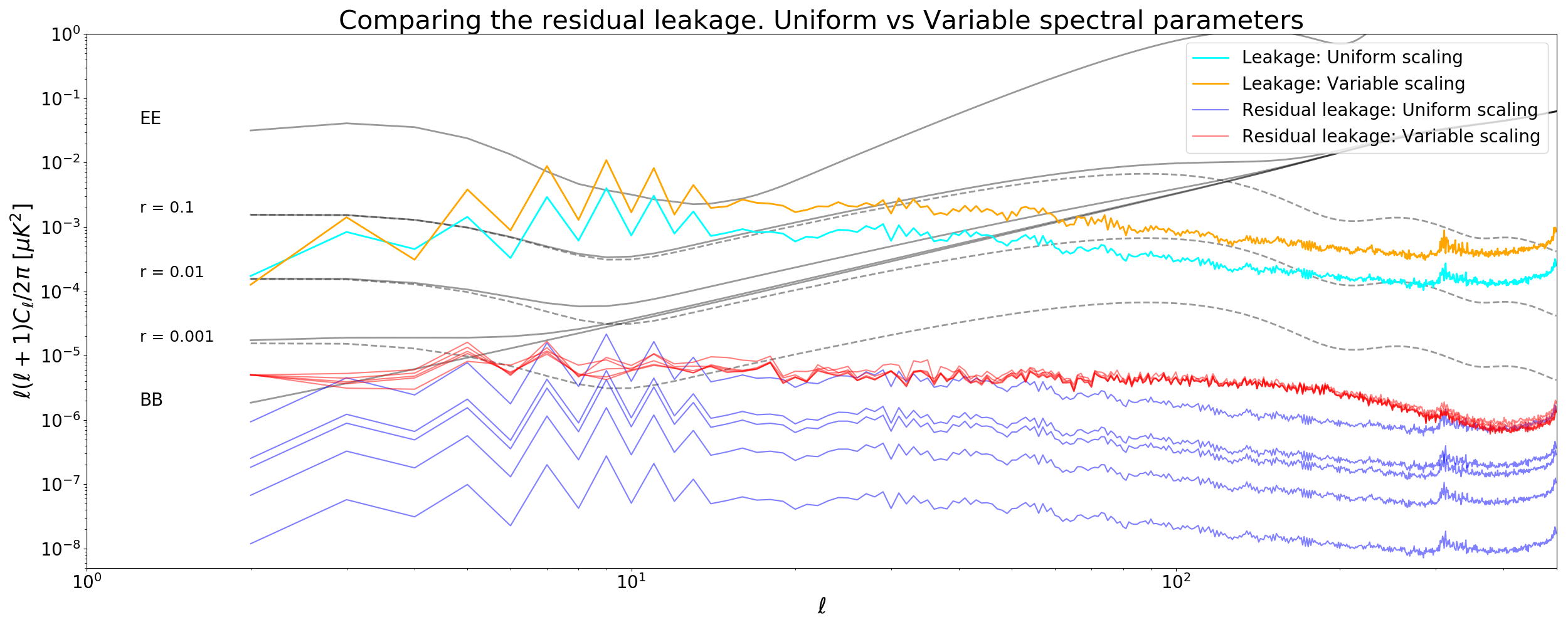}
  \caption{We demonstrate the limit of the correction algorithm due to intrinsic noise in the timestream and the template maps, and also the limit due to spatially varying spectral parameters of the foreground component. This is performed at $140\,\tr{GHz}$ and with thermal dust as the only contaminant. We construct two input cases, one where the thermal dust has a variable spectral index (scaling) and greybody temperature, and one where these are uniform. The BB spectrum of the bandpass leakage for these cases are shown in orange and cyan respectively. Both cases have noise in the timestream and in the template maps. In the latter case, with uniform scaling of the thermal dust, the residual spectra (blue) varies significantly more than the former (red) and is due to the uncertainty on estimating the amplitude of the spurious signal. For the former case, higher order leakage terms not accounted for by the correction algorithm provide a floor for the residual spectra.}
  \label{fig:uniform_vs_variable}
\end{figure}

In figure~\ref{fig:uniform_vs_variable} we see that when our input and template was uniformly scaled, the residual has a very wide scatter. This is attributed to the uncertainty in estimating the amplitude of the leakage signal. However, when our input has spatially varying spectral parameters, the residue from the map alone is dominant and masks the effect of the uncertainty in measuring the leakage amplitude.

\clearpage
\newpage
%
%
\bibliographystyle{unsrt}
\bibliography{biblio}



\end{document}